 [2005/06/22 5.2/AAS markup document class]%

\documentclass[preprint2,usenatbib,]{aastex}%
\makeindex
\makeatletter
\let\o@verbatim\verbatim
\def\verbatim{%
  \ifhmode\unskip\par\fi
  \ifx\@currsize\normalsize
     \small
  \fi
  \o@verbatim
}

\renewcommand \verbatim@font {%
  \normalfont \ttfamily
  \catcode`\<=\active
  \catcode`\>=\active
}

\RequirePackage{shortvrb}
\MakeShortVerb{\|}

\begingroup
  \catcode`\<=\active
  \catcode`\>=\active
  \gdef<{\@ifnextchar<\@lt\@meta}
  \gdef>{\@ifnextchar>\@gt\@gtr@err}
  \gdef\@meta#1>{\m{#1}}
  \gdef\@lt<{\char`\<}
  \gdef\@gt>{\char`\>}
\endgroup
\def\@gtr@err{%
   \ClassError{ltxguide}{%
      Isolated \protect>%
   }{%
      In this document class, \protect<...\protect>
      is used to indicate a parameter.\MessageBreak
      I've just found a \protect> on its own.
      Perhaps you meant to type \protect>\protect>?
   }%
}
\def\verbatim@nolig@list{\do\`\do\,\do\'\do\-}

\newcommand{\m}[1]{\mbox{\it #1}}

\def\cmd#1{\cs{\expandafter\cmd@to@cs\string#1}}
\def\cmd@to@cs#1#2{\char\number`#2\relax}
\DeclareRobustCommand\cs[1]{\texttt{\char`\\#1}}

\newcommand{\be}{\begin{equation}}
\newcommand{\ee}{\end{equation}}
\newcommand{\ba}{\begin{eqnarray}}
\newcommand{\ea}{\end{eqnarray}}

\newcommand{\degree}{\ensuremath{\mathrm{^\circ}}}
\newcommand{\arcm}{\ensuremath{\mathrm{^\prime}\;}}
\newcommand{\arcs}{\ensuremath{\arcmm\hskip -0.1em\arcmm \;}}
\newcommand{\arcmm}{\ensuremath{\mathrm{^\prime}}}

\newcommand{\dotsec}{\,\rlap{\hbox{$\mathrm{^s}$}}{\hbox{$.$}}\,}

\received{}
\revised{}
\accepted{}

\ccc{ASP}

\usepackage{graphicx}

\shorttitle{}
\shortauthors{A. Bonafede et al.}

\begin{document}

\title{%
Evidence for particle re-acceleration in the radio relic in the galaxy cluster PLCKG287.0 +32.9
}%

\author{A. Bonafede\altaffilmark{1},  H. T. Intema\altaffilmark{2}, M. Br{\"u}ggen\altaffilmark{1},} 
\affil{$^1$ Hamburger Sternwarte, Universit\"at Hamburg, Gojenbergsweg 112, 21029, Hamburg, Germany. \\}
\affil{$^2$ National Radio Astronomy Observatory, 1003 Lopezville Road, Socorro, NM 87801-0387, USA\\}

\author{M. Girardi\altaffilmark{3,4}, M. Nonino\altaffilmark{4},}
\affil{$^3$ Dipartimento di Fisica-Sezione di Astronomia, Universit\'a di Trieste, via Tiepolo 11, I-34143 Trieste, Italy\\}
\affil{$^4$ INAF - Osservatorio Astronomico di Trieste, via Tiepolo 11, I-34143 Trieste, Italy\\}

\author{N. Kantharia\altaffilmark{5}, R. J. van Weeren\altaffilmark{6}, H. J. A. R{\"o}ttgering.\altaffilmark{7}}
\affil{$^5$ National center for Radio Astrophysics, TIFR, Post Bag 3, Ganeshkhind, Pune 411 007, India}
\affil{$^6$Harvard-Smithsonian Center for Astrophysics, 60 Garden Street, Cambridge, MA 02138, USA}
\affil{$^7$Leiden Observatory, Leiden University, 2300 RA Leiden, the Netherlands}

\begin{abstract}

Radio relics are diffuse radio sources observed in galaxy clusters, probably produced by shock acceleration 
during cluster-cluster mergers. Their large size, of the order of 1 Mpc, indicates that the emitting
electrons need to be (re)accelerated locally. The usually invoked Diffusive Shock Acceleration models
have been challenged by recent observations and theory.
We report the discovery of complex radio emission in the Galaxy cluster 
PLCKG287.0 +32.9, which hosts two relics, a radio halo, and several radio filamentary emission. 
Optical observations suggest
that the cluster is elongated, likely  along an intergalactic filament, and displays a significant amount of substructure.
The peculiar features of this radio relic are that (i) it appears to be connected to the lobes of a radio galaxy and (ii) the radio spectrum steepens on either side of the radio relic. We discuss the origins of these features in the context of particle re-acceleration.

\end{abstract}

\keywords{Galaxy clusters, non-thermal processes}

\maketitle
%\tableofcontents

\section{Introduction}

The hierarchical model of structure formation predicts that clusters
of galaxies grow by successive accretion of smaller sub-units. A large
amount of energy is dissipated in the intra-cluster medium (ICM) as a
result of merger events, and a fraction of this energy could be
channeled into the amplification of the magnetic fields \citep[see e.g.][for a review]{Dolag08}
 and into the acceleration of high-energy Cosmic
Rays \citep[see][for a review]{Petrosian08} that in turn may
produce observable extended radio emission. Extended radio sources on a cluster
scale, not associated with any optical counterpart, have been detected
in a number of galaxy clusters \citep{Venturi07, Giovannini09,2011A&A...533A..35V,Bonafede12,Feretti12}.
They are called radio halos and radio relics, depending on their morphology,
location and radio properties.\\ 

\noindent {\bf Radio relics}\\ 

Radio relics are extended sources, characterised by a steep spectrum\footnote{The spectrum is here defined as $S(\nu) \propto \nu^{-\alpha}$.} ($\alpha >$1) and
strong polarisation ($\sim$ 20-30 \% at 1.4 GHz). Their origin is
subject to debate and not yet understood. There is a general consensus that they
are related to shock waves, occurring in the ICM during mergers. Shock
waves should be able to amplify magnetic fields and accelerate
electrons up to relativistic energies, hence producing synchrotron
radio emission  \citep{Bruggen11,2012MNRAS.423.2781I,Vazza12,sk13}. Although this
picture is roughly consistent with observations, some important issues remain unexplained. 
First, a radio relic is not always detected when a shock wave is present in the ICM, as revealed by X-ray observations 
\citep{Russell11} and recently \citet{2013MNRAS.433..812O} have found a displacement
between the X-ray shock wave and the radio relic in 1RXS J0603.3+4214. Secondly, most of the
relics do not appear to be co-located with a shock wave in X-ray observations (see review by \citealt{Bruggen11} and ref. therein). 
Thirdly, shock waves in
the ICM are characterised by Mach numbers of the order of 2 - 4 \citep{sk08,va09shocks}
 so that their efficiency
in accelerating particles is expected to be too low to account for the radio emission.\\
Some authors have proposed that
shock waves by themselves are not sufficient to accelerate the particles from the thermal pool to 
relativistic energies. This would indicate that a seed population of relativistic electrons must be
already present before the shock passage \citep{KangRyu11,KangRyu12,Pinzke13}.
 This seed population of old radio plasma could be
re-energised by a shock wave, explaining why the connection between a
relic and a shock wave is not one to one. Since the old radio plasma is
much more energetic than the thermal gas in the ICM, a low
acceleration efficiency could be sufficient to re-accelerate the particles
and power the radio emission.  
The problem is now to find the source for this seed 
population of relativistic electrons. Some authors have suggested that they come
 from a previous episode of shock acceleration (\citealt{Macario11}).
Another possibility is that the old plasma comes from the activity of radio galaxies.\\

Recent results by \citet{VazzaBruggen13} have shown that the gamma-ray upper limits to the
cluster  emission from the {\it Fermi} satellite are at odds with the hypothesis that radio relics 
are generated by Diffusive Shock Acceleration (DSA) mechanism, at least in the way it appears to work for supernova remnants. In fact, if shock waves accelerate electrons in the ICM, also protons should be 
accelerated, with a higher efficiency. In inelastic collisions with the thermal protons, relativistic protons produce  pions and gamma-ray photons, which {\it Fermi} should have detected.\\

\noindent {\bf Radio Halos}\\

Radio halos are diffuse Mpc-sized objects found in merging galaxy clusters \citep[e.g.][]{Buote01,Cassano10}
and characterised by steep spectra (see e.g. reviews by \citealt{Ferrari08,Feretti12}).
Two different classes of models have been proposed so far: the hadronic models \citep{2010ApJ...722..737K,2011A&A...527A..99E}
and the turbulent re-acceleration models \citep{Fujita03,CassanoBrunetti05}. 
In this paper, we study the peculiar and complex  emission of the cluster 
PLCKG287.0 +32.9 to test  DSA models.
PLCKG287.0 +32.9  is a massive cluster ($M_{500} \sim 1.4 \times 10^{15}$ solar masses, Planck Collaboration 2013)
located at redshift $z=0.39$, showing a disturbed X-ray morphology
 (Bagchi et al. 2011) which indicates that the cluster is undergoing a merger
 event. The X-ray centre of the cluster has a Right  Ascension equal to  
 $11h50m51.02s$ and a declination of  $-28d04'09.37"$
  PLCKG287.0 +32.9 is known to host two radio relics  \citep{Bagchi11}.
Our new radio observations give a different picture of the system.
The radio emission is more complex and more extended than previously observed, and offers a unique opportunity to
unravel the origins of radio relics.
The paper is organised as follows: In Sec.  \ref{radio} 
we present our new observations. The X-ray  and optical data are discussed in Sec. \ref{sec:xray} and \ref{sec:optical}, respectively.
In Sec. \ref{radioanalysis} the radio properties are studied, and the spectral properties are presented in Sec. \ref{sec:spix}. Results are discussed in \ref{discussion} 
and we present our conclusions in Sec. \ref{conclusions}.\\
Throughout this paper, we assume a
concordance $\rm{\Lambda CDM}$ cosmological model, with $H_0=$ 71 km
s$^{-1}$ Mpc$^{-1}$, $\Omega_M=$ 0.27, and $\Omega_{\Lambda}=$
0.73. One arcmin corresponds to 316 kpc at $z=$0.39.

\begin{figure}

\vspace{90pt}
\begin{picture}(90,90)

%\put(-20,0.20){\includegraphics[width=9.5cm]{Xray_radioHR.eps}}
%\put(250,10){  \includegraphics[width=8cm]{label.eps}}

 \put(-20,0){\includegraphics[width=9cm]{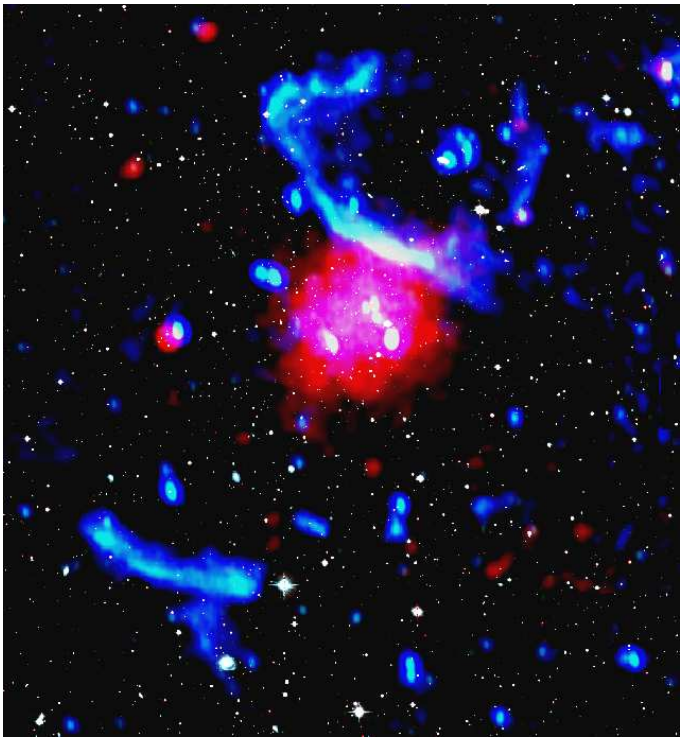}}     %{rgb2.eps}}
 \end{picture}
 \caption{X-ray emission in red (XMM-Newton), radio emission at 323 MHz in blue (low resolution, beam FWHM $\sim 22'' \times 18''$) and in green (high resolution:  beam FWHM$\sim 13'' \times 8''$).}
\label{fig:xray}
\end{figure}

\begin{table*}
 \centering
   \caption{Radio observations.}
  \begin{tabular}{c c c c c c c c}
  \hline
  Frequency & Radio telescope & Observing date &  time  & bandwidth  & Int. time & rms noise & Beam \\
   MHz        &                &                         &    h        &    MHz          &    s                 & mJy/beam &\\
\hline
325   &  GMRT    &    Jan 2013 & 	 7        & 33             &    8 		  & 0.1               & 13$'' \times 8''$    \\  
610   &  GMRT    &    May 2013&         10     & 33              &    8		   & 0.065          & 7$'' \times 5''$ \\
150    &  GMRT    &	Jul  2011 &        8     & 16              &    8		   & 1.3               &25$'' \times 18''$  \\
 3000   &  JVLA      &	Jun 2013 &         6	  & 2000         &    3                 & 0.05    & 19$'' \times 11''$  \\
\hline

\multicolumn{7}{l}{\scriptsize  Col. 1: Observing  Frequency, Col. 2: Radio telescope. Col. 3: Date of the observation. Col. 4: total observing time. }\\
\multicolumn{7}{l}{\scriptsize  Col. 5: Observing bandwidth. Col 6: Integration time per visibility.}\\
\multicolumn{7}{l}{\scriptsize Col 7: 1$\sigma$ rms noise reached in the high resolution images.}\\ 
\multicolumn{7}{l}{\scriptsize Col 8: FWHM of major and minor axes of the restoring beam in the high resolution images. }\\
\end{tabular}
\label{tab:obs}
\end{table*}

\section{Radio observations}
\label{radio}
\subsection{GMRT observations}
The cluster was observed at the Giant Meterwave Radio Telescope (GMRT) at 325 MHz, 610 MHz and 150 MHz (archival data).
In Table \ref{tab:obs} we list the details of the radio observations.
At 323 MHz the observations were recorded using a 33 MHz bandwidth subdivided into 256 channels and
8s integration time.  At 610 MHz we used a bandwidth of  33 MHz subdivided in 512 channels and 8s integration
time. At 610 MHz the observations were split into two separate observing blocks due to scheduling constraints.
We retrieved from the archive observations at 150 MHz. These observations were recorded in one sideband 16 MHz wide, split into 128 channels,
and using 8s integration time.

The source 3C147 was observed for 15 minutes at the beginning of the observing block, and used as absolute flux and 
bandpass calibrator at 150 and 325 MHz, adopting a flux of 64.8 Jy and 53.1 Jy at 150 MHz and   323 MHz, respectively \citep{ScaifeHeald12}.
3C286 was used as a bandpass and absolute flux calibrator for the 610 MHz observations. Following \citet{ScaifeHeald12}
we adopted a flux of 22.3 Jy.
3C147   was also used to estimate the instrumental contribution to the
antenna gains, which is also needed for ionospheric calibration at 323 MHz  and 150 MHz (see below).
The amplitude and bandpass gains were then  directly transferred to the target field. At 150 MHz and 323 MHz the instrumental phase 
information was also used to correct the target field.\\
Strong Radio Frequency Interferences (RFI) was removed from the target field data by statistical outlier flagging tools. Much of the remaining low-level RFI 
was modelled and subtracted from the data using  Obit \citep{OBIT}.
 After RFI removal, datasets have been averaged down to 28 (150 MHz) 24 (323 MHz) and 28 (610 MHz)
channels in order to speed up the imaging process and to avoid significant bandwidth smearing.\\
For the phase calibration, we started from a model derived from the NRAO Vla Sky Survey (NVSS, \citealt{NVSS})
 and then proceeded with self-calibration loops. We decided not to use a phase calibrator, as the GMRT field of view is wide, and a non-negligible 
flux  is present in the field of the available  phase calibrators.
We note that using a phase calibrator  to bootstrap the flux from 3C147/3C286 would have altered the flux scale, leading to higher errors in the flux measurements. 
The imaging  has been performed using AIPS. In order to compensate for the non-coplanarity of the array we used wide-field imaging technique, decomposing the GMRT field of view into $\sim$ 100 facets. 
 We performed rounds of cleaning and self-calibration, inspecting the residual visibilities for a more accurate removal of low-level RFIs. 
In order to correct for ionospheric effects, leading to direction-dependent phase errors, we applied SPAM calibration \citep{SPAM} and imaging to the target field at 323 and 150 MHz. 
The presence of strong sources in the field-of-view enabled us to derive directional-dependent gains for each of them (similarly to the peeling technique) and to use these gains to fit a phase-screen over the entire field of view.
The final images of the full field-of-view centered on the target were  corrected for the primary beam response. We have used the 
 408 MHz all-sky map by Halsam et at. (1995) to correct for system temperature variations between the calibrator and the target field.
The final images are shown in Fig. \ref{fig:radio}.
The highest resolution images are made using the Briggs weighting scheme (robust=-1 in the AIPS definition), and including all the available 
range of visibilities in the uv-plane (Figs. \ref{fig:label} and \ref{fig:radio}). As a result, we should be more sensitive to the
large-scale emission in the low-frequency bands.\\
We estimate that the residual amplitude errors are of the order of 6\% at 610 and 323 MHz, and $\sim 10\%$ at 150 MHz, in line with values
reported for GMRT observation at these frequencies \citep[e.g.][]{2009A&A...506.1083V,Intema11,Bonafede12,Macario13}.\\

\subsection{VLA observations}
\begin{figure*}
\vspace{200pt}
\begin{picture}(200,200)
\put(230,0){      \includegraphics[width=8cm]{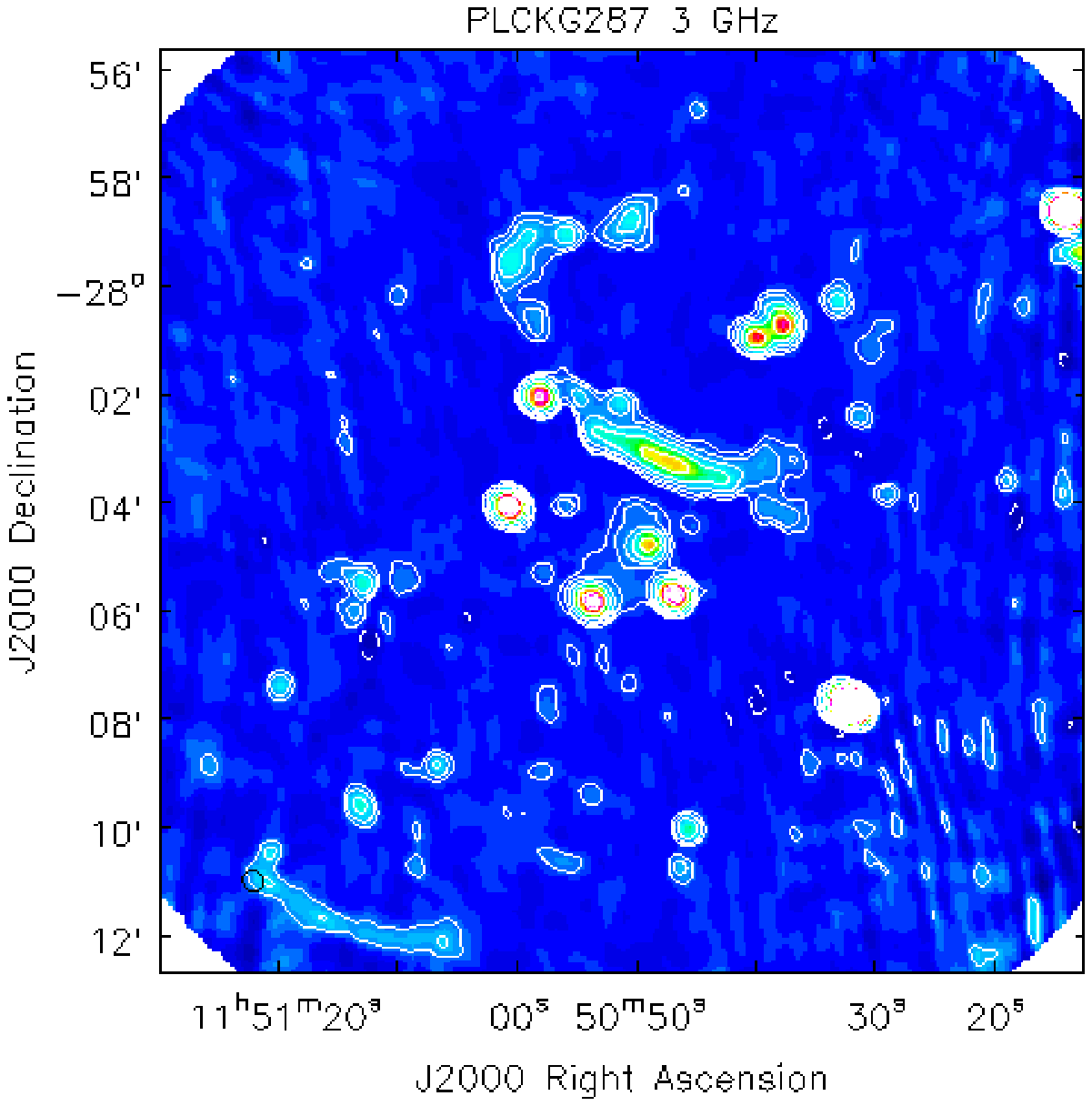}}
\put(-25,0){    \includegraphics[width=8.2cm]{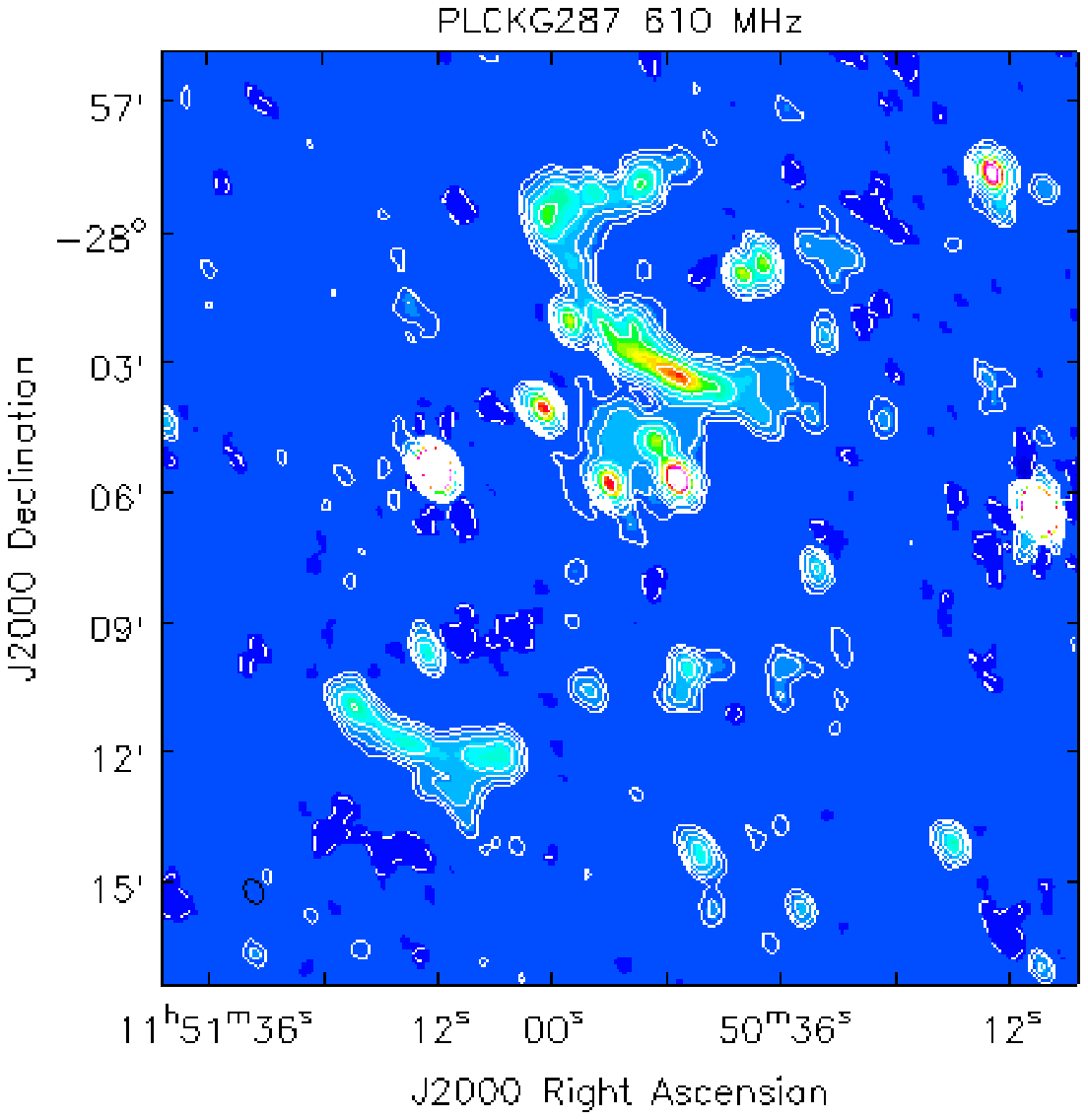}}
\put(230,250){ \includegraphics[width=8cm]{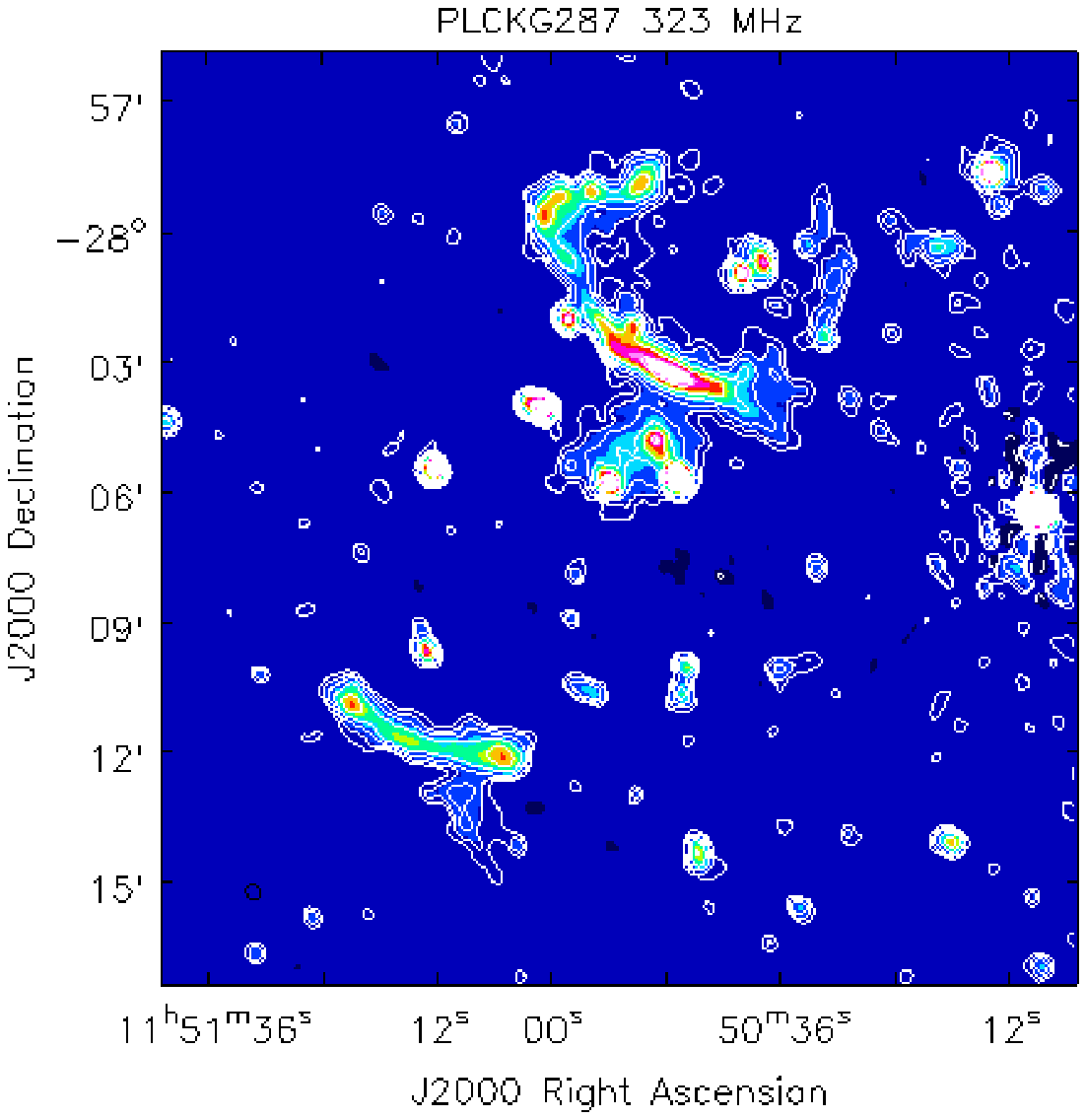}}
\put(-25,250){\includegraphics[width=8.5cm]{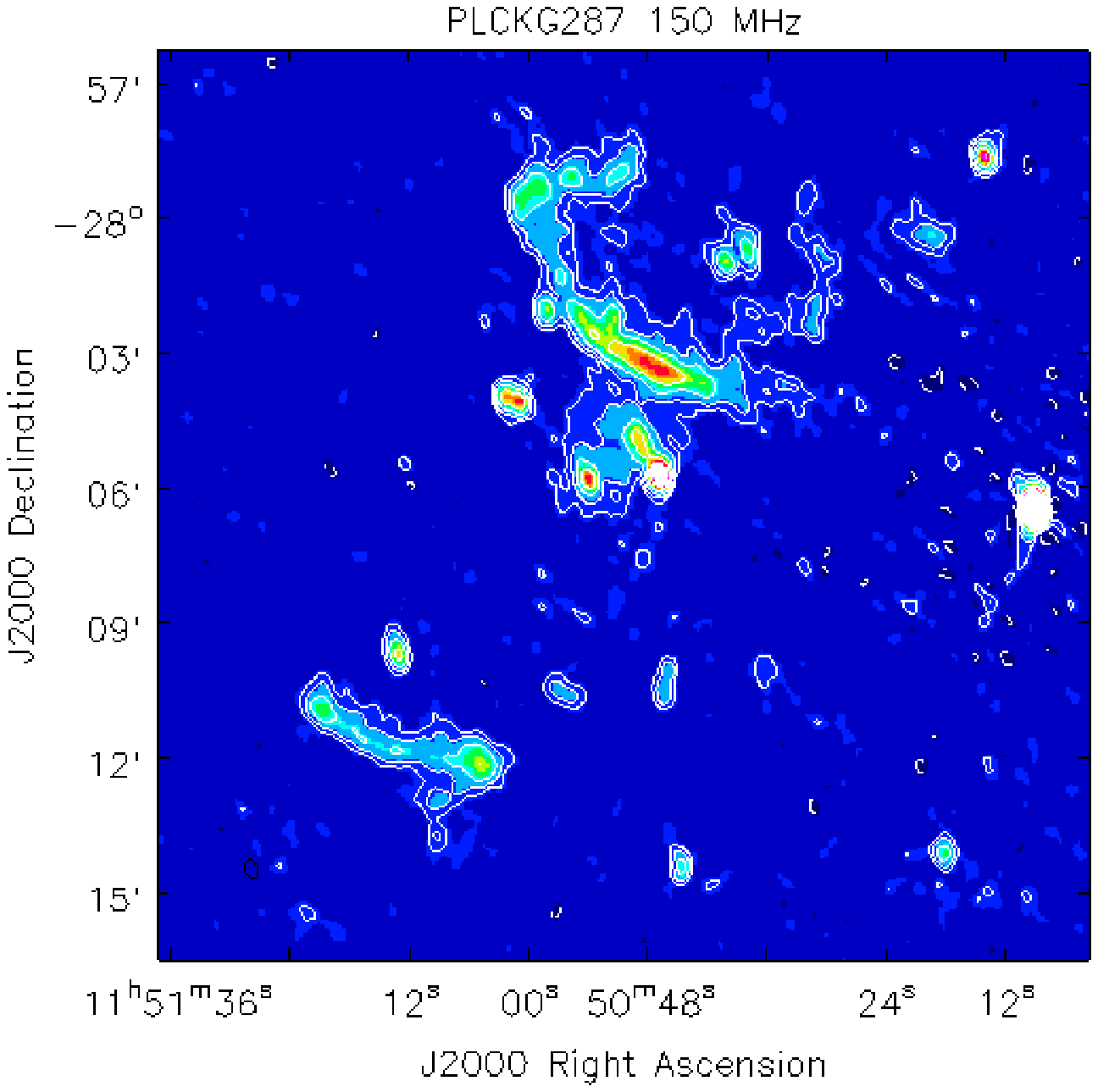}}

\end{picture}
 \caption{Radio emission from PLCKG287.0 +32.9 at 150 MHz (top, left), 323 MHz (top, right), 610 MHz (bottom left), and 3 GHz (bottom right.)
 The image at 150 is obtained with a Briggs weighting scheme, intermediate between uniform and natural weighting, to lower the side-lobes of the primary beam.
 The image at 323 MHz is obtained using  a Gaussian taper to enhance the diffuse emission. The images at  610 MHz and  3 GHz are obtained 
 cutting the longer baselines and using a Gaussian taper. Contours start at $\pm 3 \sigma$ and are spaced by a factor 2.
 The 1$\sigma$ values are are 1.3 mJy/beam (150 MHz), 0.15 mJy/beam (323 MHz),  0.13 mJy/beam (610 MHz), and 30 $\mu$Jy/beam (3 GHz).}
\label{fig:radio}

\end{figure*}

 Karl G. Jansky Very Large Array (VLA) observations of the cluster were performed in C band (2-4 GHz) in 
 the DnC to C array configuration transition period.
Due to scheduling constraints the observations were conducted in three separate
scheduling blocks of 2h each. 
The two base-bands were centered on frequencies of 2.5 GHz and 3.5 GHz. Both have a  bandwidth of 1 GHz, covering 2 to 4 GHz continuously.
Every baseband is divided into 9 spectral windows, subdivided into 64 channel each, with a frequency resolution
of 2 MHz/channel.  The integration time was 3s per visibility. 
This set-up gives a high frequency and time resolution, which is needed to carefully remove the RFIs.
 However, RFI is severe in the 2-4 GHz band, especially at low elevation.
  Only 11 spectral windows could be used for imaging and  almost 40\% of the time has been flagged because
 of RFIs. The net bandwidth is  1.4 GHz. Details on the observations are listed in Table \ref{tab:obs}.\\
Calibration and imaging were performed using the Common Astronomy Software Applications (CASA 4.2).
We observed  3C147 as bandpass and absolute flux calibrator, and we used  the \citet{PerleyButler13} flux scale,
which is the most accurate at frequencies above 1 GHz.  
The source J1146-2447 was observed every $\sim$30 min and it was used as a phase calibrator.
We used the multi-frequency multi-scale implementation of the CASA clean to account for the wide bandwidth and the 
different angular scales of the radio emission. Specifically, we used 2 terms of the Taylor expansion to model the frequency
dependence of the sky emission.\\
We used the peeling technique to derive direction-dependent gains towards a bright source in the field and subtracted it.
This bright source has a strongly inverted spectrum ($\alpha \sim -2$) which would complicate the imaging step.
Cycles of phase only self-calibration were performed to refine the antenna-phase gain variations on the target field. 
The residual amplitude errors are estimated to be $\sim6\%$.
The final image was then corrected for the primary beam attenuation, computed for each spectral window separately to account for the
wide bandwidth (Fig. \ref{fig:radio}).

\section{X-ray observations}
\label{sec:xray}

From the XMM-Newton data archive, we have downloaded a 10 ksec observation of the cluster.
The X-ray image is shown in Fig. \ref{fig:xray}. The emission is disturbed and slightly 
elongated in the NW-SE direction.  The X-ray morphology confirms the dynamically perturbed state of the cluster.
The Planck collaboration (2011) reports  a temperature $T_x $ of 13 keV, and an X-ray luminosity 
$L_{(0.1-2.4 keV)} \sim 1.72 \times 10^{45} \rm{erg s^{-1}}$ within $r_{500}$.

\begin{figure*}

\vspace{230pt}
\begin{picture}(230,230)

\put(10,380){\includegraphics[width=5.5cm]{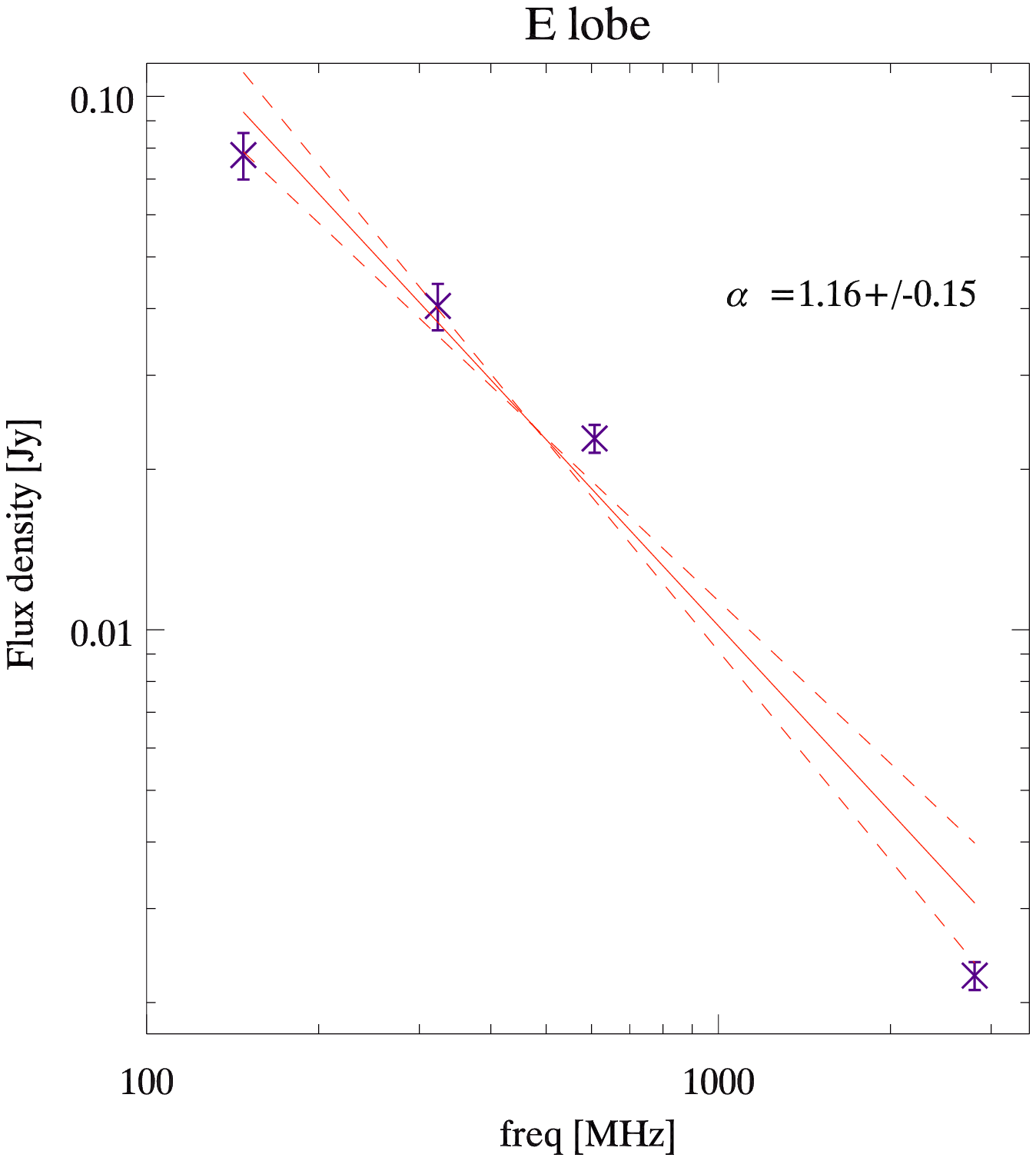}}
\put(170,390){\includegraphics[width=5.5cm]{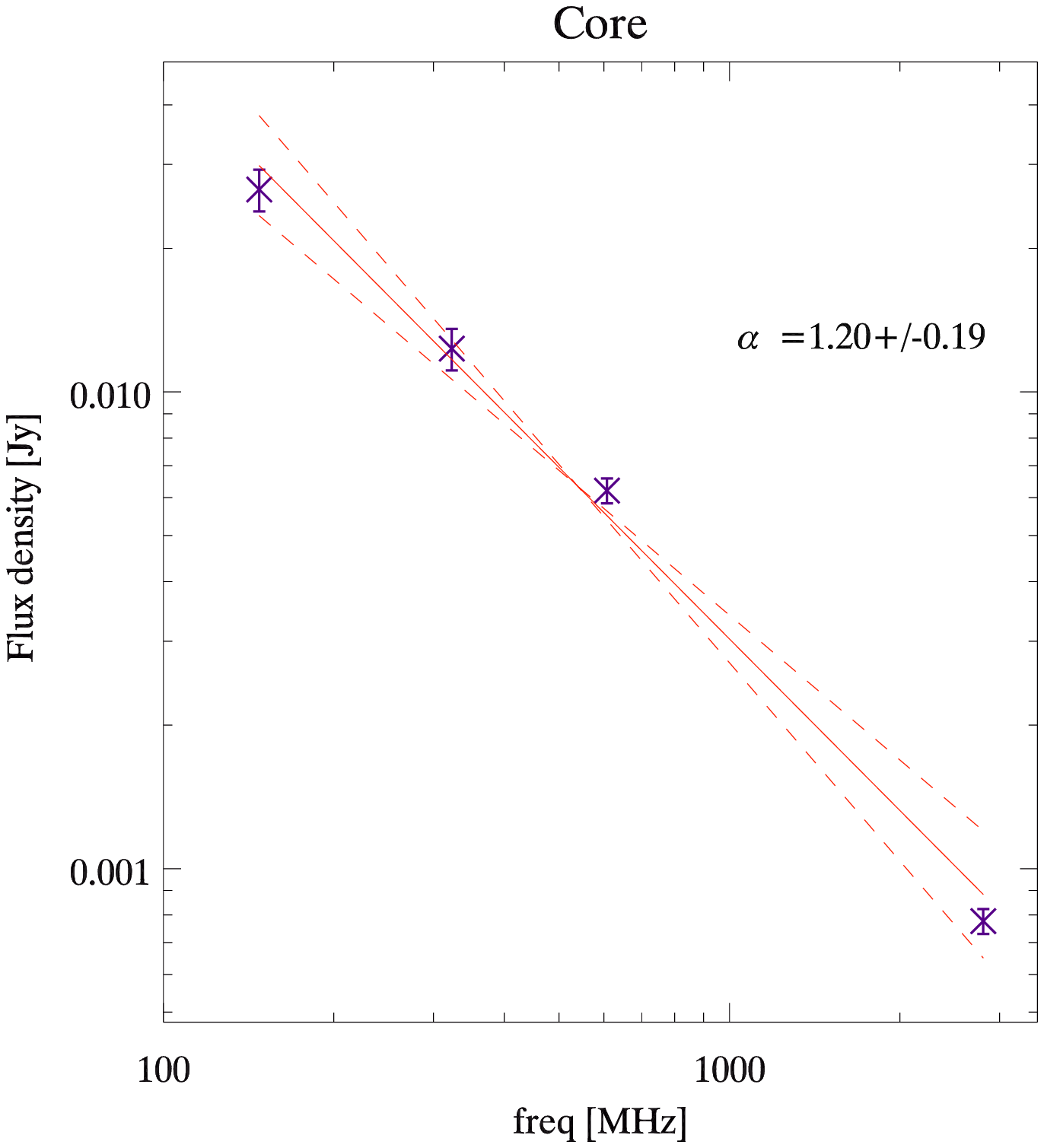}}
\put(330,380){\includegraphics[width=5.5cm]{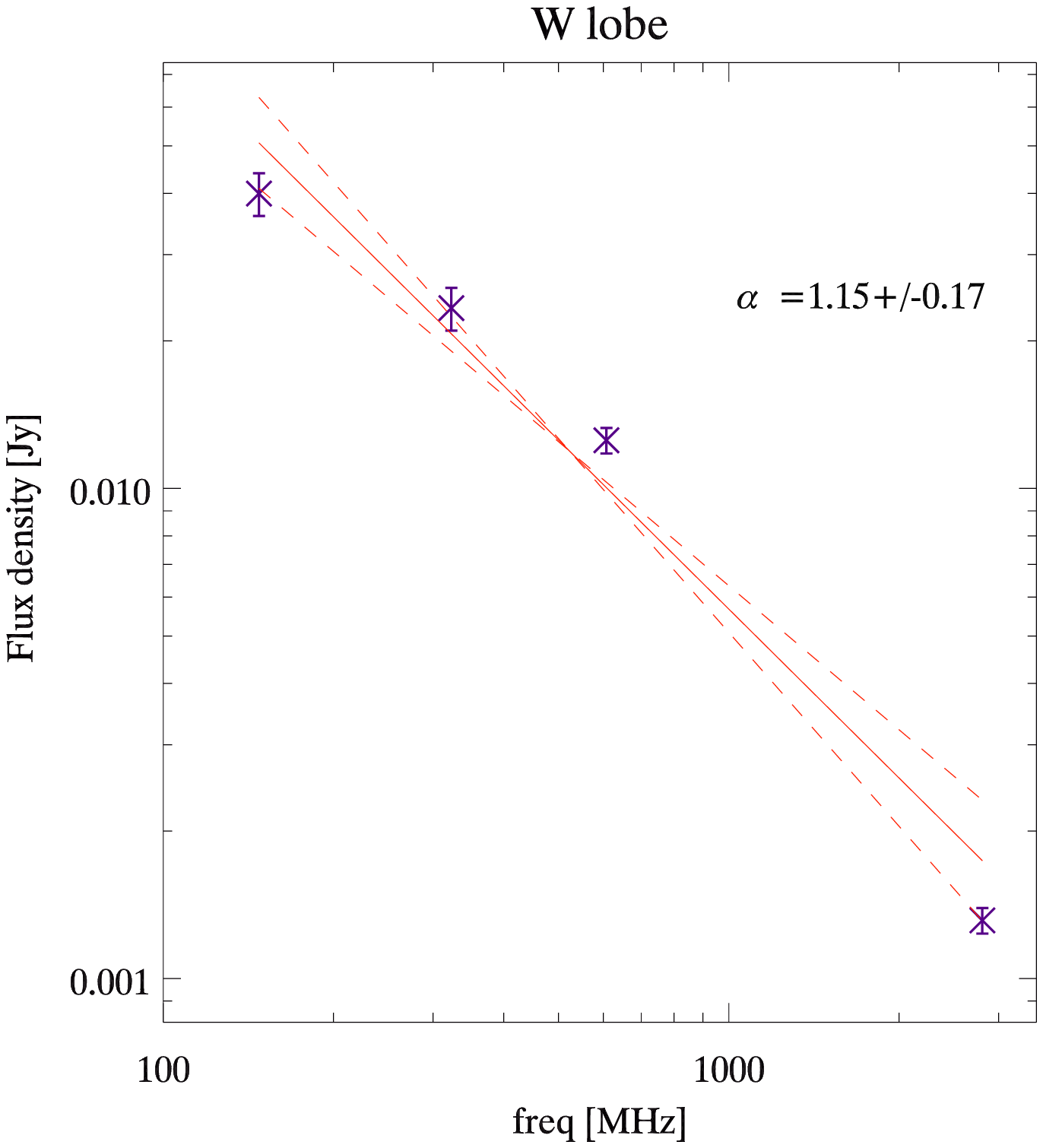}}

\put(-20,200){\includegraphics[width=5.5cm]{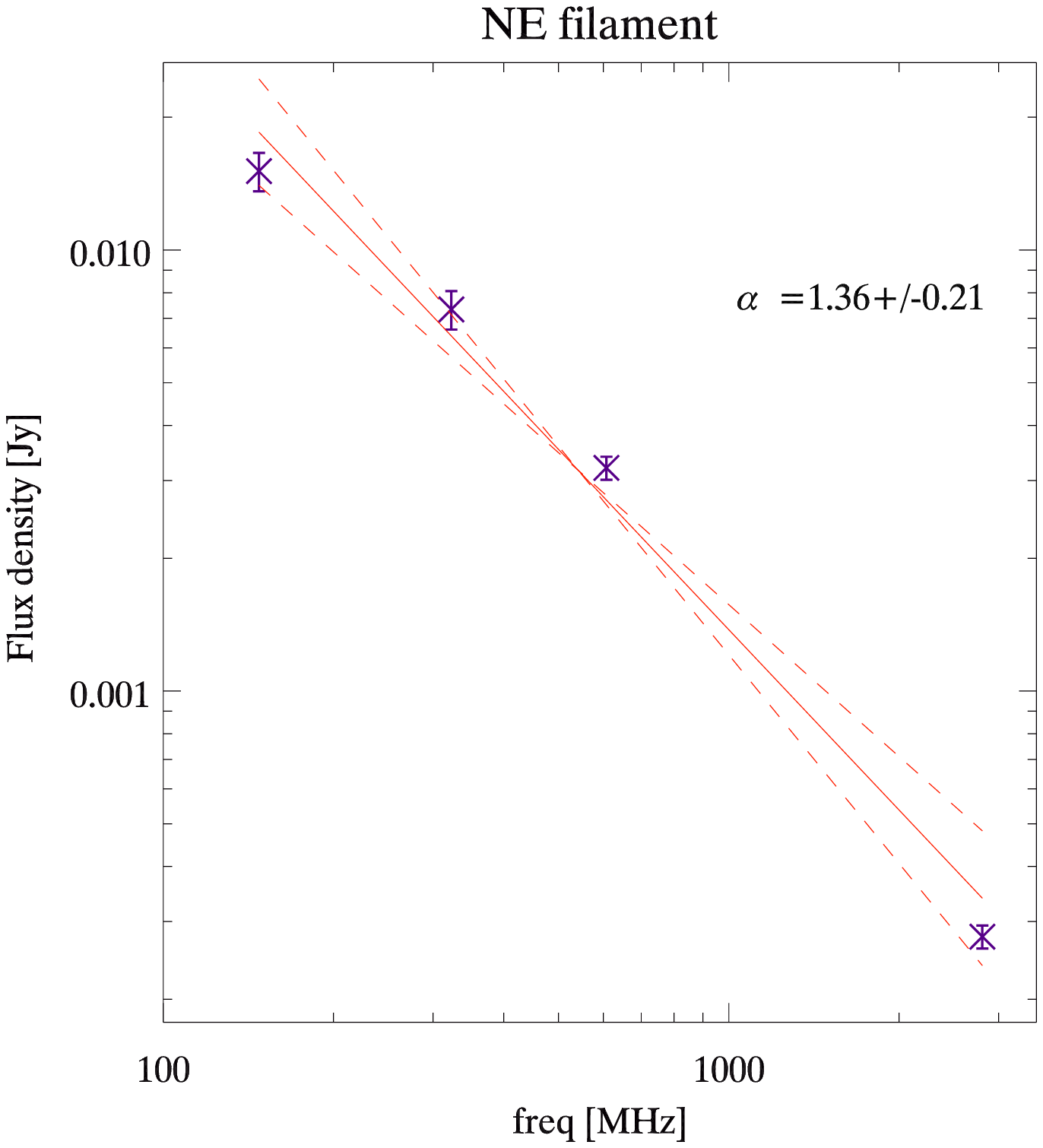}}

\put(10,20){\includegraphics[width=5.5cm]{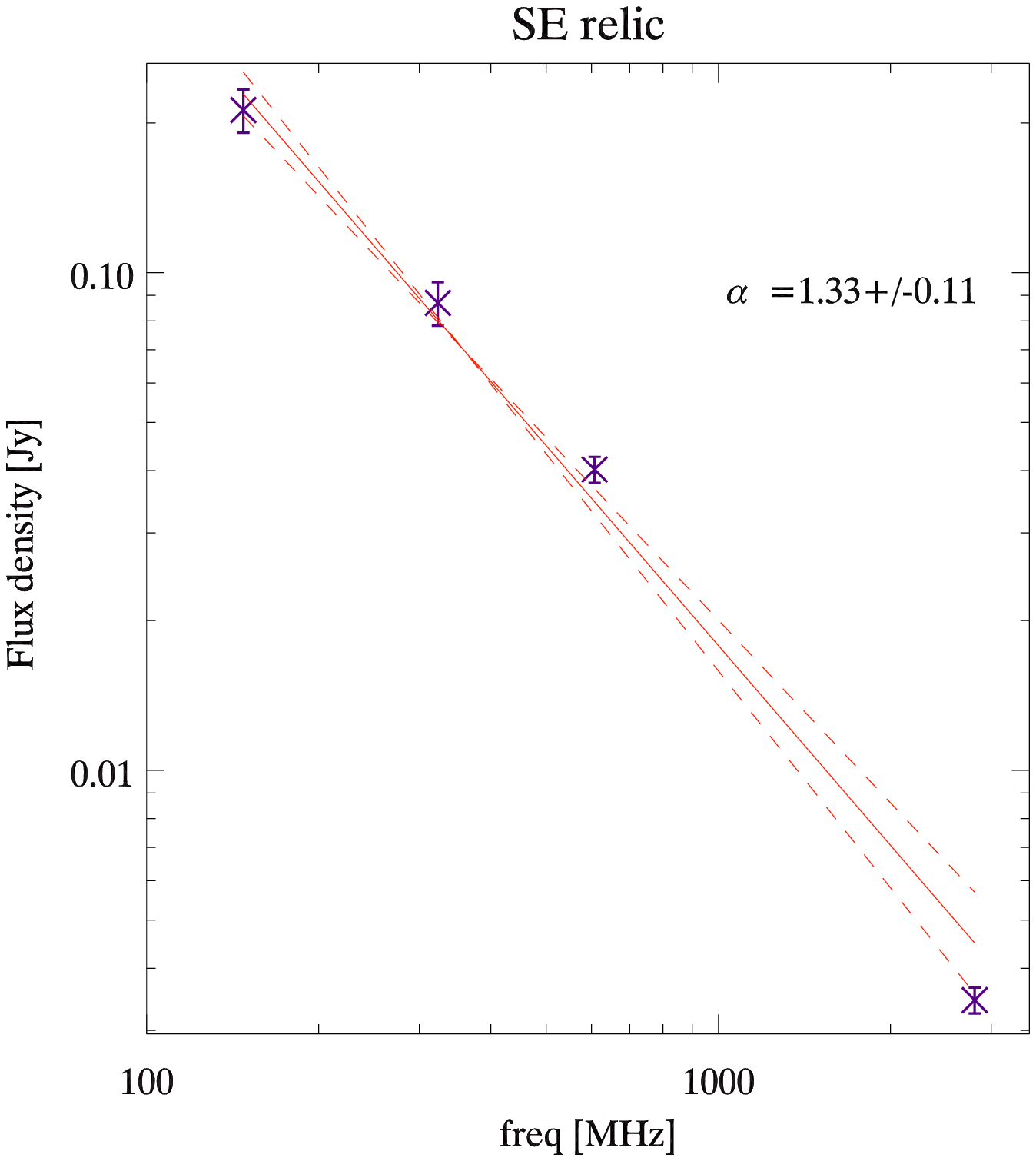}}
\put(170,10){\includegraphics[width=5.5cm]{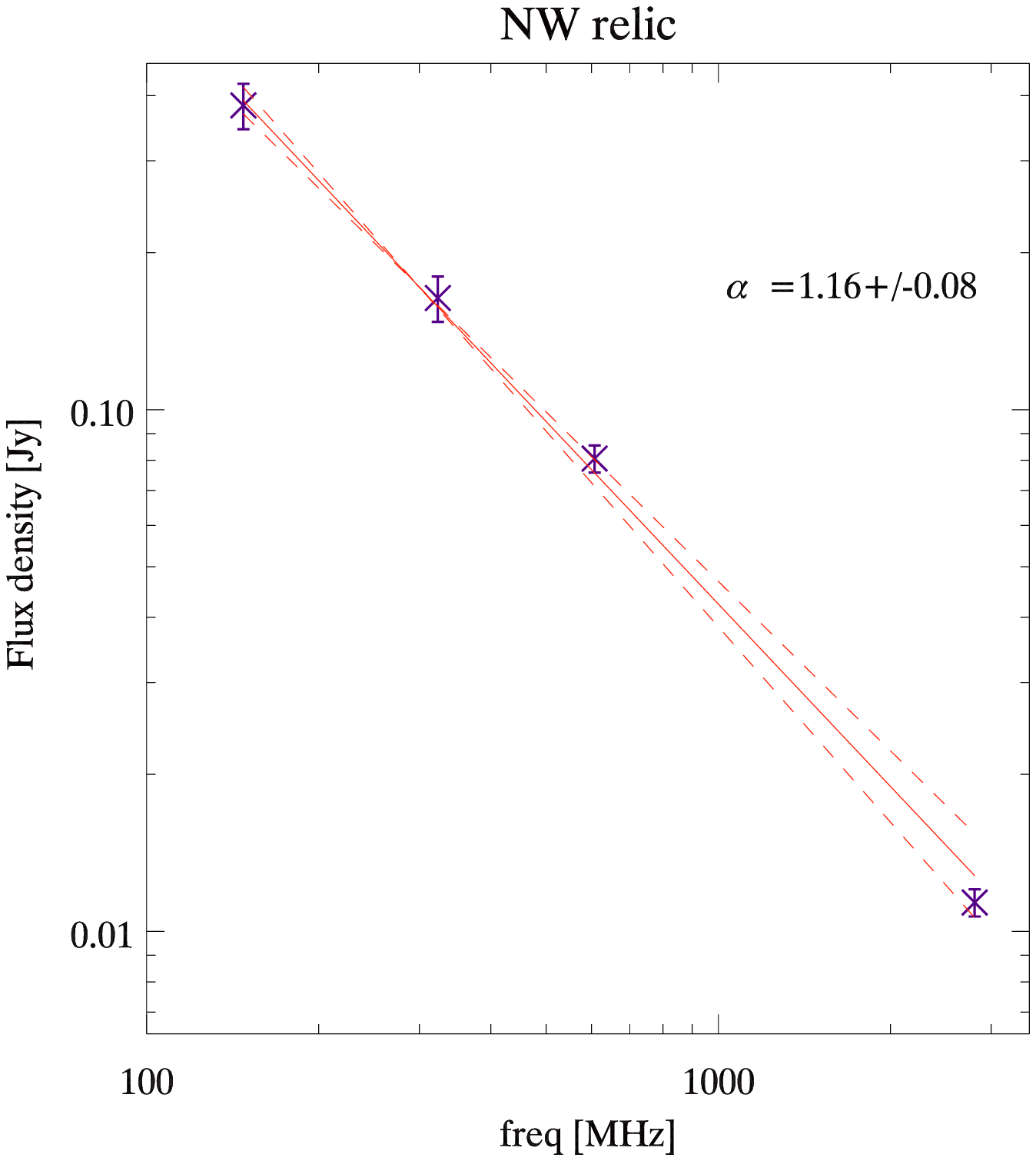}}
\put(330,20){\includegraphics[width=5.5cm]{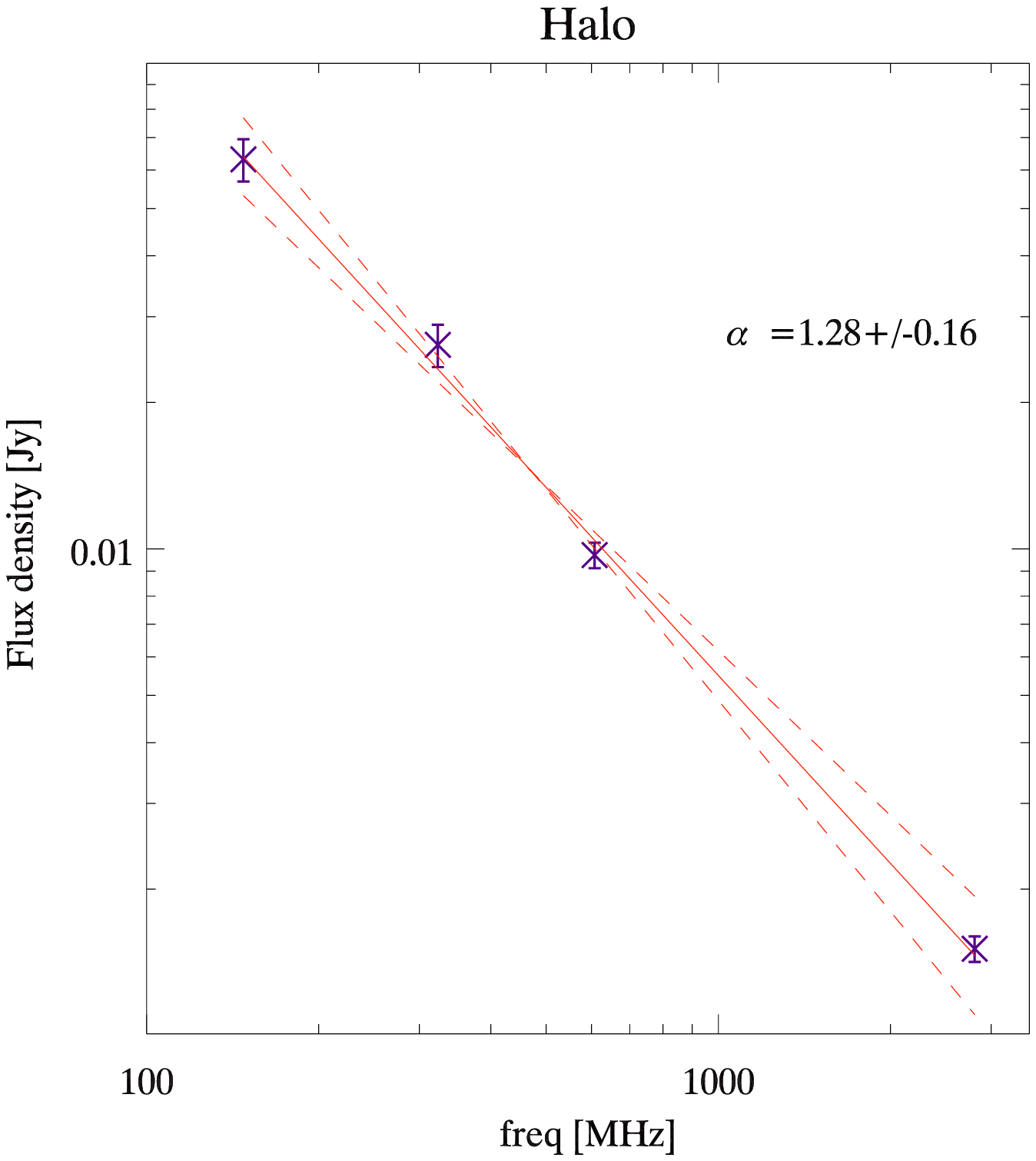}}

\put(360,200){\includegraphics[width=5.5cm]{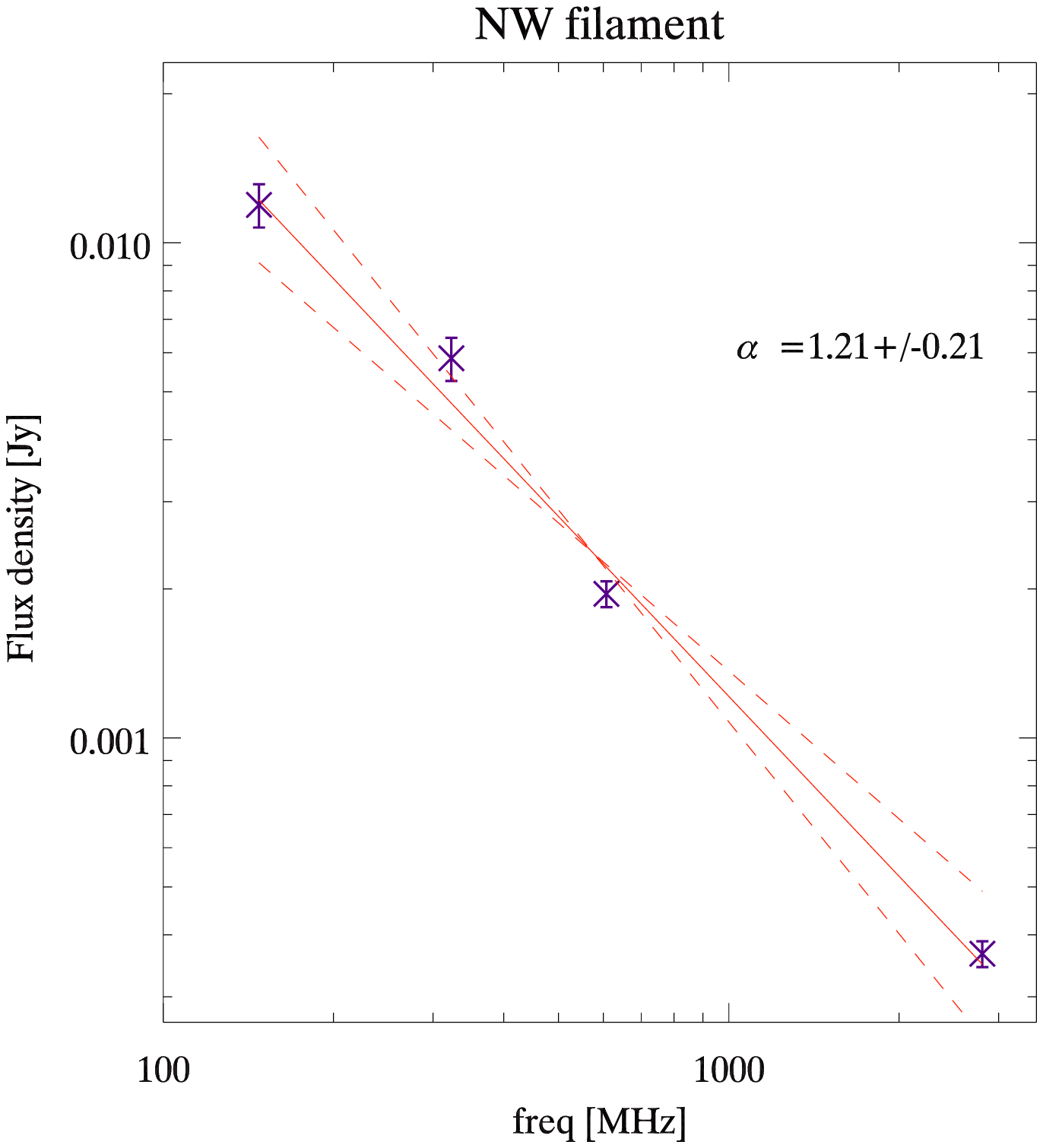}}

\put(140,170){\includegraphics[width=8cm]{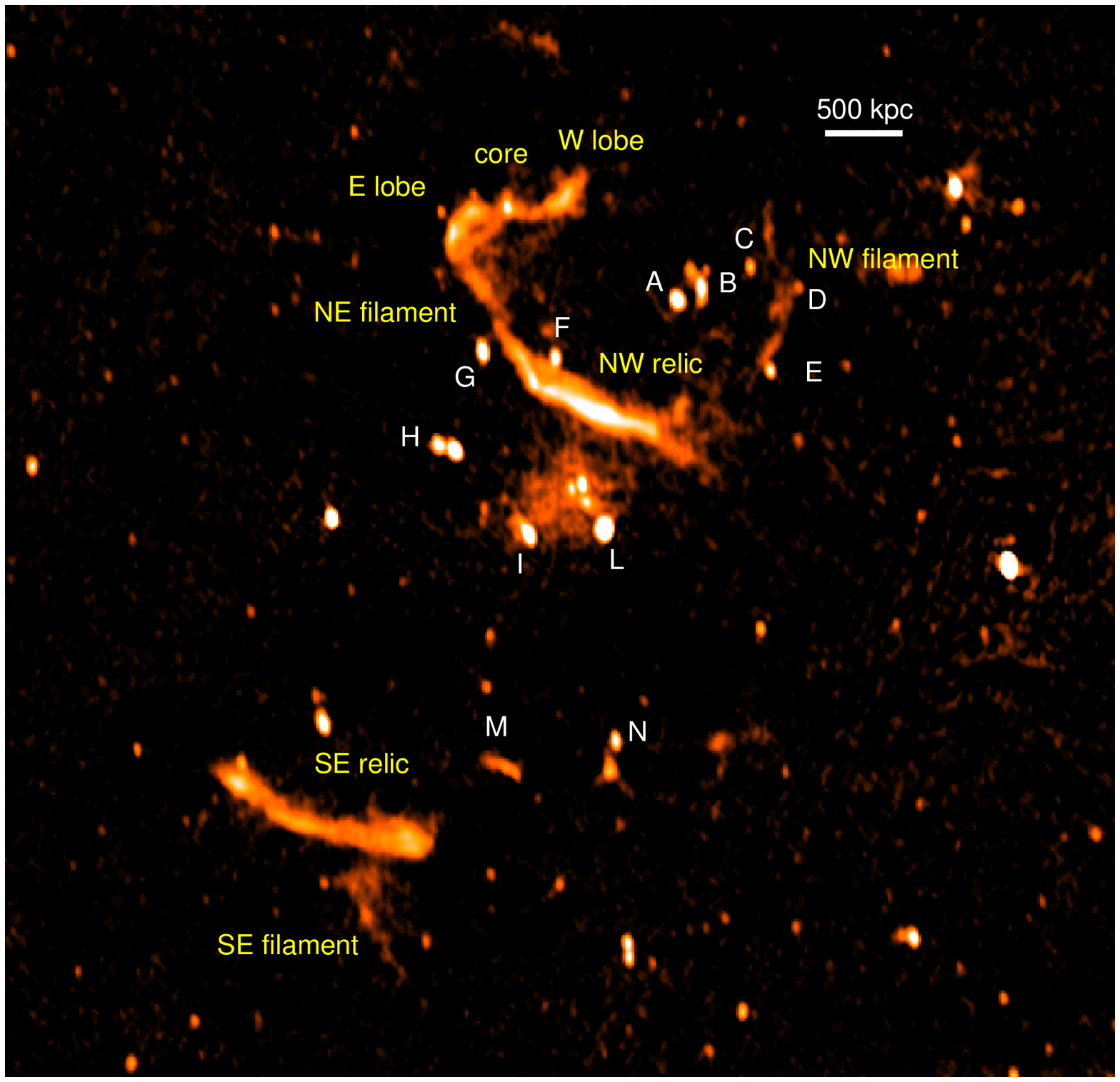}}

\end{picture}
\caption{Central panel: Radio emission at 323 MHz. Yellow labels indicate the diffuse sources detected in the cluster field. White labels mark the embedded radio-galaxies. 
The surrounding panels show the spectral index fits for the extended emission in the cluster. The continuous line shows the attempt to fit a power-law from 150 MHz to 3 GHz, 
the dashed lines are the fit 1$\sigma$ errors. Note that the flux has been measured only where the 
emission is detected above 3$\sigma$ at all frequencies.}
\label{fig:label}
\end{figure*}

\begin{figure}
\centering
%\resizebox{\hsize}{!}{\includegraphics{figcm.eps}}
\includegraphics[width=8cm]{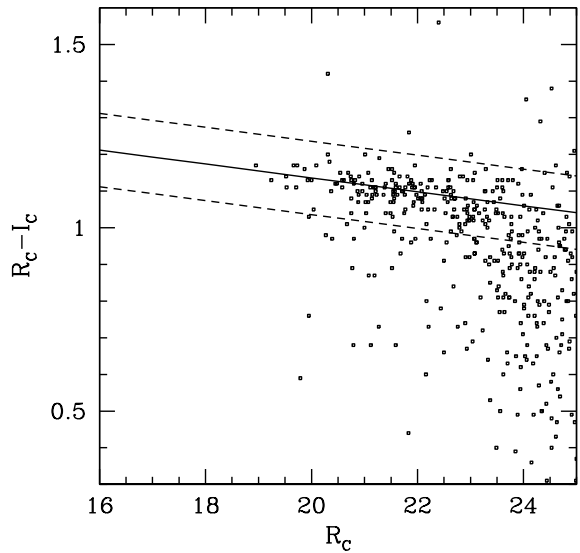}
\caption
{WFI $R_{\rm C}$--$I$ vs. $R_{\rm C}$ diagram for galaxies within the
  0.5 Mpc cluster region available.  The solid line gives the CMR
  obtained from the fit and the dashed lines are drawn at $\pm$0.1 mag to
  indicate the region used to define the likely member galaxies in the
  entire photometric catalog.}
\label{figcm}
\end{figure}

\begin{figure*}

\vspace{100pt}
\begin{picture}(100,100)
\put(-15,0){  \includegraphics[width=9cm]{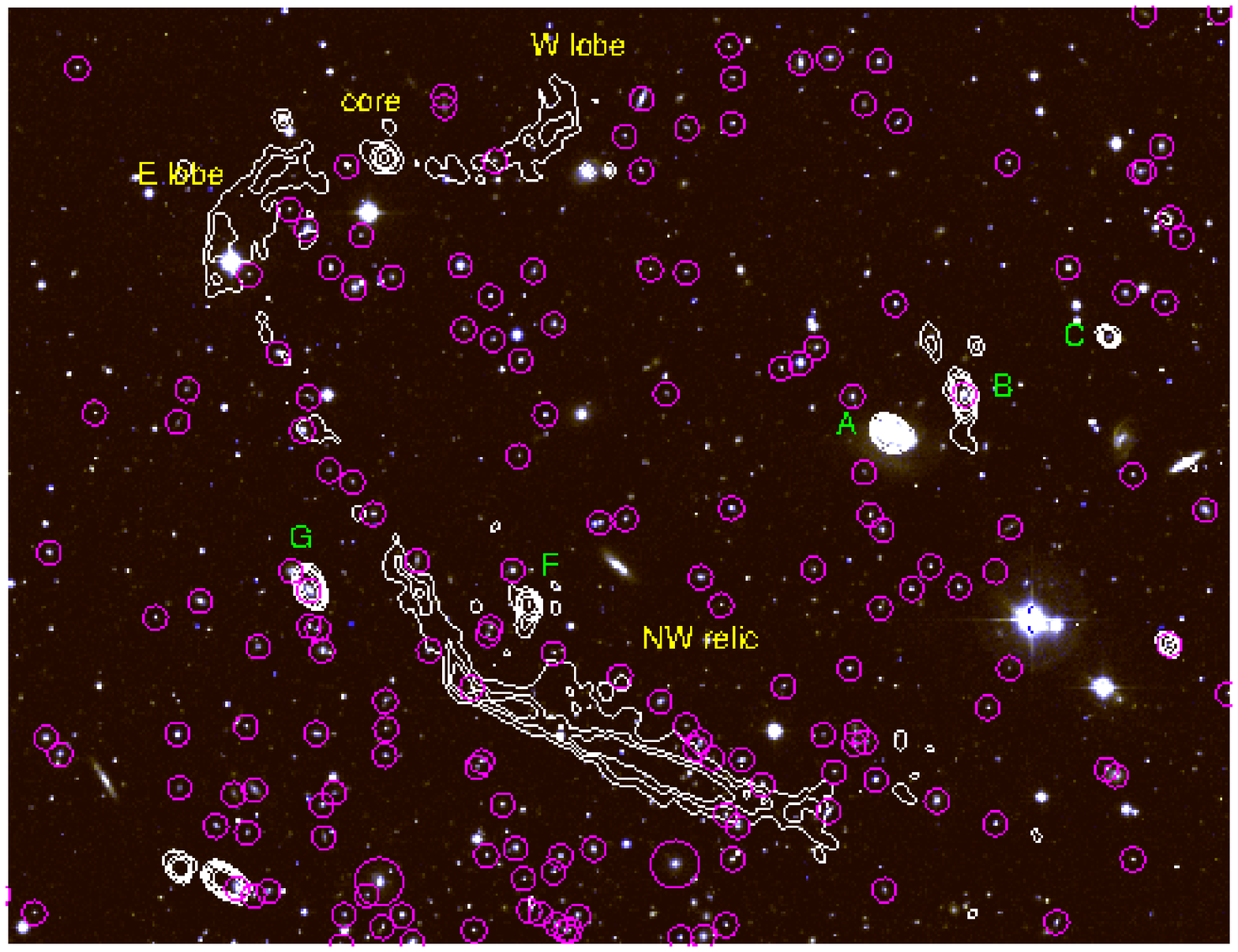}}
 \put(235,12){  \includegraphics[width=8.5cm]{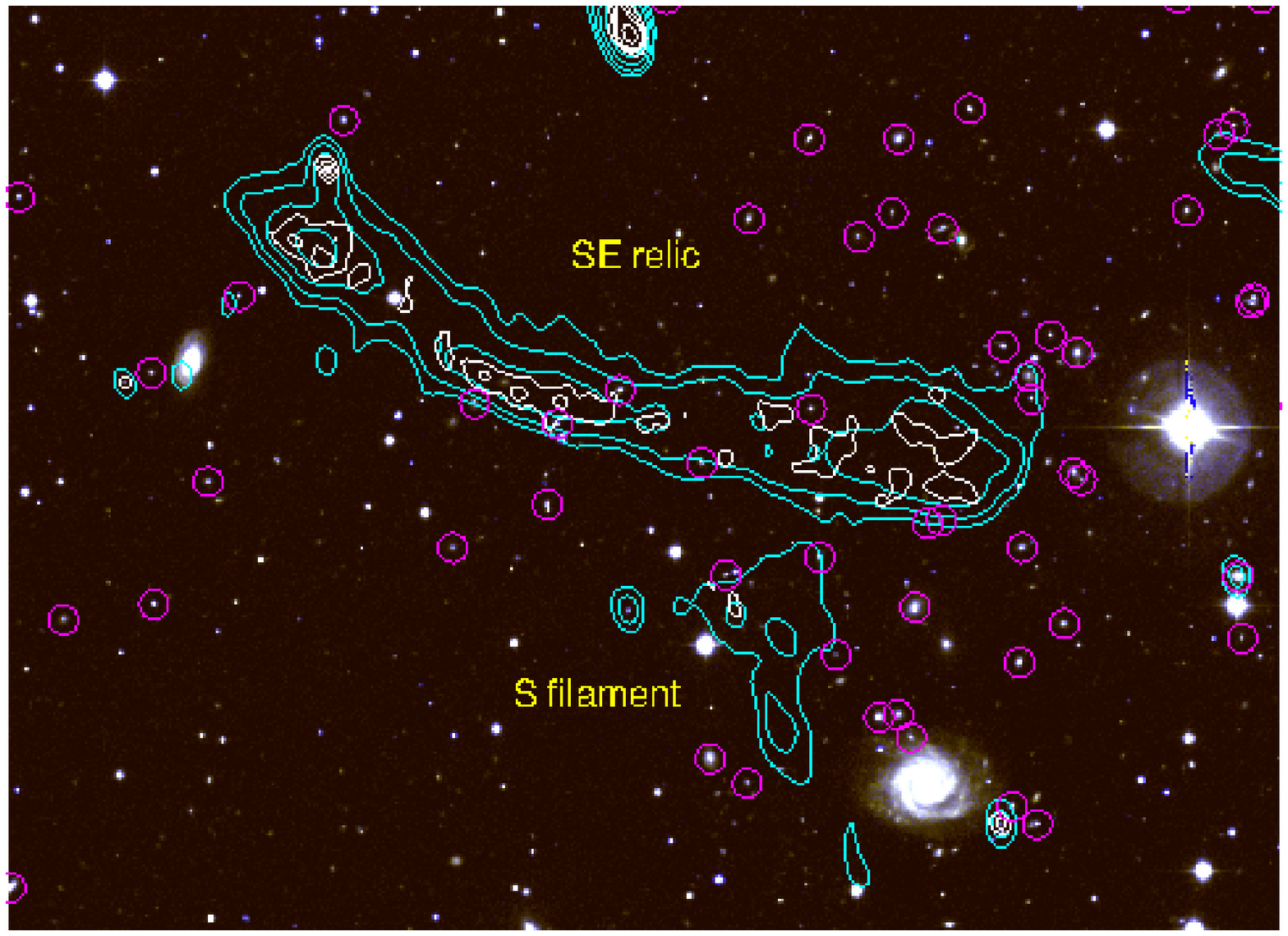}}
 \end{picture}
 \caption{The NW (left) and SE relic (right): optical emission in colours.
  Magenta  circles indicate the likely members with
	$R<22$, white contours are from 610 MHz high-resolution observations. 
	The beam is $\sim 7'' \times5''$, contours start at 3$\sigma$ (0.195 mJy/beam) and are spaced by
	a factor 2. In order to highlight the SE relic, 323 MHz contours are plotted in cyan.
	The beam is $\sim 13''\times8''$, contours start at 3$\sigma$ (0.3 mJy/beam) and are spaced by
	a factor 2. Labels indicate the radio sources (as in Figure \ref{fig:label}.)}
\label{fig:optradio}
\end{figure*}

\begin{figure}

\vspace{100pt}
\begin{picture}(100,100)
\put(-10,0.4){\includegraphics[width=9cm]{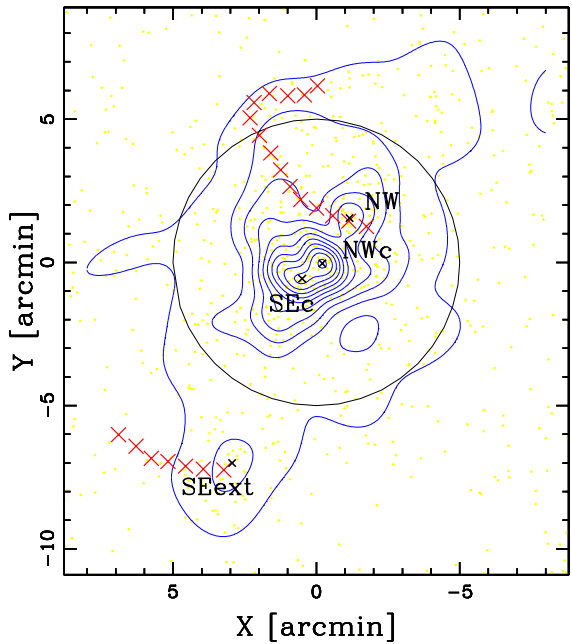}}
 \end{picture}
 \caption{Spatial distribution on the sky and relative isodensity
  contour map of the photometric likely cluster members with $R_{\rm
    C} \le 22.5$, obtained with the 2D-DEDICA method.  The BCG
  position is taken as the cluster center. The circle indicates the
  $5\arcmin$ region, i.e. $\sim 1.6$ Mpc at the cluster redshift, and
  corresponding to $\sim R_{500}$. Large red crosses indicate, in a
  schematic way, the southern and northern relic features. Small black
  crosses and labels indicate the density peaks listed in
  Table~\ref{tabdedica2d}. }
\label{fig:isodensity}
\end{figure}

\begin{table}
        \caption[]{2D substructure from the WFI photometric sample.}
         \label{tabdedica2d}
            $$
         \begin{array}{l r c c r }
            \hline
            \noalign{\smallskip}
            \hline
            \noalign{\smallskip}
\mathrm{Subclump} & N_{\rm S} & \alpha({\rm J}2000),\,\delta({\rm J}2000)&\rho_{
\rm S}&\chi^2_{\rm S}\\
& & \mathrm{h:m:s,\degree:\arcmm:\arcs}&&\\
         \hline
         \noalign{\smallskip}
\mathrm{NWc}         & 101&11\ 50\ 49.2-28\ 04\ 58&1.00&68\\
\mathrm{SEc}         & 230&11\ 50\ 52.4-28\ 05\ 30&0.92&78\\
\mathrm{NW}          &  92&11\ 50\ 44.9-28\ 03\ 23&0.52&46\\
\mathrm{SEext}       &  91&11\ 51\ 03.5-28\ 11\ 56&0.20&26\\
              \noalign{\smallskip}
              \noalign{\smallskip}
            \hline
            \noalign{\smallskip}
            \hline
         \end{array}
$$
%\begin{list}{}{}
%\item[$^{\mathrm{a}}$] Corresponding to the galaxy cluster SDSS C.
%\item[$^{\mathrm{b}}$] Corresponding to the galaxy cluster SDSS CE
%\end{list}
\end{table}

\section{Optical analysis}
\label{sec:optical}
ESO 2.2 m Wide Field Imager (WFI) images\footnote{Based on
  observations made with ESO Telescopes at the La Silla Paranal
  Observatory under programme ID 086.A-9028(B).} were retrieved from
the ESO Science Archive. Following the same steps as adopted by \citet{non09}, the images in V-band, R$_{\rm C}$-band, and
I$_{\rm C}$-band were added and reduced to cover about a
30\arcmm$\times$30\arcm field of view.  In summary, our stacks were
based on a total exposure time of 5400 s in V-band (1.7\arcsec
seeing), 10\,800 s in R$_{\rm C}$-band (1.0\arcsec), and 6000 s in
I$_{\rm C}$-band (0.8\arcsec).  The astrometric solutions were
obtained using 2MASS as reference.  Unfortunately, the observations in
all filters were collected under non-photometric conditions, as shown
from the relative photometric analysis. Therefore, the final stacked
images were not calibrated with standard stars.  For the color selection, we used 2MASS to make a first-order calibration. The
absolute zero-point calibration was estimated to be accurate to within
$0.2$ mag. We obtained a photometric catalog of $\sim$ 50\,000 objects
using SExtractor  \citep{ber96}.

After a preliminary rejection of stellar objects using the flux radius
Sextractor parameter, we selected likely members on the basis of the
($R_{\rm C}$--$I$ vs. $R_{\rm C}$) color-magnitude relation (hereafter
CMR).  Due to the limitations in the $V$-band data, we preferred to not
use them to improve the member selection.  In order to determine the CMR, we
applied the 2$\sigma$-clipping fitting procedure to the galaxies
within 0.5 Mpc of the cluster center.  For the center of PLCK287+32.9,
we adopted the position of the BCG
[R.A.=$11^{\mathrm{h}}50^{\mathrm{m}}50\dotsec1$, Dec.=$-28\degree
  04\arcmm 55.3\arcs$ (J2000.0)]. The resulting CMR is $R_{\rm
  C}$--$I$=$1.52(\pm 0.07)-0.019(\pm 0.03)\times$ $R_{\rm C}$ (see
Fig.~\ref{figcm}).  Out of our photometric catalog, we considered
{\it likely} cluster members to be those objects lying within 0.1 of the
CMR. Density reconstruction of the 2D galaxy distribution was performed
through the 2D-DEDICA adaptive kernel method \citep{pis93}.  
The cluster structure is
elongated along the NW-SE direction and exhibits significant
substructure both in central and external regions
($R_{500}\footnote{The radius $R_{\delta}$ is the radius of a sphere
  with mass overdensity $\delta$ times the critical density at the
  redshift of the galaxy system.}\sim 1.6$ Mpc; Planck Collaboration, 2011). 
  In particular, Figure~\ref{fig:isodensity} shows the contour map
for the likely cluster members having R$_{\rm C}\le 22.5$, thus
sampling the galaxy luminosity function down to $\sim 1.5$ mag fainter
than R$_{\rm C}^*$ (1473/633 galaxies in the whole
field/$R_{200}$-region).  Table~\ref{tabdedica2d} lists information
for the four most significant ($>>$ 99\% confidence level)  density peaks: the
number of assigned members, $N_{\rm S}$ (Col.~2); the peak position
(Col.~3); the density (relative to the densest peak), $\rho_{\rm S}$
(Col.~4); the value of $\chi^2$ for each peak, $\chi ^2_{\rm S}$
(Col.~5).  These main features are quite robust with respect to changes in the
magnitude limit.  In the very central region the main feature is the
presence of two comparable very dense peaks at NW and SE
(hereafter NWc and SEc), the former coinciding with the BCG. In
the external region, the main feature is the presence of a SE peak
(hereafter SEext) at $\sim R_{200}$.  The corresponding SEext group is
clearly a minor substructure comprising $\sim 20\%$ ($\sim 15\%$) of galaxies contained within $R_{500}$
($R_{200}$).\footnote{See also its low density with respect to the two main
peaks SEc and NWc as listed in Table~\ref{tabdedica2d}.}  As for the
direction of the cluster elongation, the position angle of the galaxy
distribution (here measured counter--clock--wise from north) becomes
larger when considering a more extended cluster region out to $\sim
R_{500}$: from PA $\sim 120$ degrees at 0.5$R_{500}$ to PA $\sim 160$
degrees at $\sim R_{500}$.

\begin{figure}

\vspace{100pt}
\begin{picture}(100,100)
\put(-40,0){\includegraphics[width=10cm]{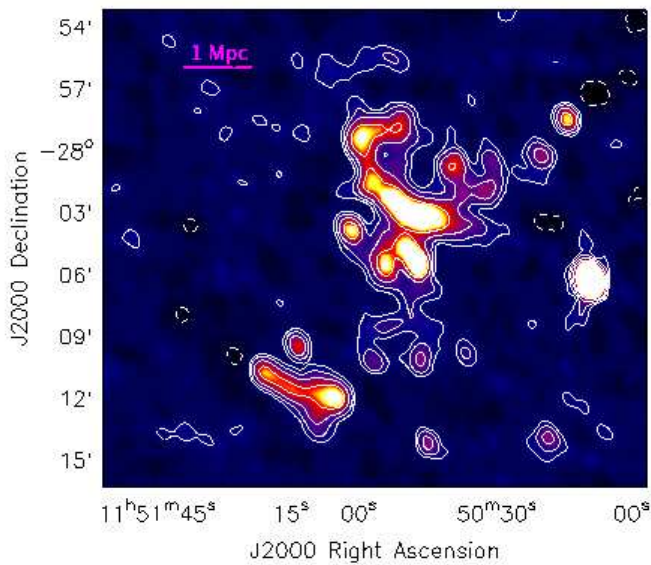}}
 \end{picture}
 \caption{Radio emission at 150 MHz in colours. The image is obtained tapering down the baselines longer than 5 k$\lambda$ to
 enhance the diffuse emission. The beam is $\sim 50" \times 40"$, the rms noise is $\sim$2 mJy/beam. Contours are drawn at
 $\pm$3$\sigma$ and are spaced by a factor 2. }
\label{fig:lr}
\end{figure}

\begin{table*}
\label{tab:radioProp}
 \centering
   \caption{Properties of the cluster radio sources.}
  \begin{tabular}{c c c c c}
  \hline
 Source         & Frequency  &  Flux density    &  LAS &  LLS  \\
 name            &  MHz         &   mJy                 &$"$     & kpc                      \\
\hline
NW relic        &  150           &   620             & 494   &  2598				              \\  %9.366642e-01
		       &  323		 &  216 		& 420   &   2209           \\%3.189992e-01
		       &  610           &   110    		& 440    &  2314  \\ %1.282870e-01
		       &3000          &	15.2   		&  360 &   1894 \\%1.691176e-02
 
 NE filament & 150             &   52    		&70    &	 368	\\	
		     &  323            &  18   		& 70  &    368              \\
		     &  610            &   7.6      		&  70 &    368  \\
		     &	3000     &  0.62      		&  60 &   316   \\

NW filament  &  150         &  84$^{*}$    & 240 & 1262    \\          %1.362584e-01   \\
		       &  323           &   14.5           &  235  &  1236             \\%3.132014e-02
		       &  610           &   5.4              &  130 &   684 \\
		       &	3000    &  0.43$^{** }$  &  80 &   421 \\
		       
lobeW         & 150        &   75   & 95  &    500 \\
		     &  323          &  33       &100  &   526 \\ 
		     &  610          &  33      & 138 &   725 \\
		     &	3000    & 1.7         & 120 &    631 \\
		     
lobeE          & 150           & 95   & 100  &  526\\
		     &  323          &   45   &  100 & 526              \\
		     &  610          &    31    & 100 & 526 \\
		     &	3000   &    2.7     & 80  & 421 \\
		     
core                &  150     &   30    & -  & - \\
			 & 325       &    13.9      & -  & -\\
		       &  610       &    3.9       &  -  & -\\
		       &	3000  &    0.65        &  -  & -   \\
		       &&&&\\
		       
Halo	     &  150       & 314  &  450 &    2104 \\ %9.406868e-01 
		     &   323       &  63 &  250 &    1315 \\%2.457847e-01               \\
		     &   610       &  26    &  250 &    1315\\             %1.026415e-01 
		     &  3000	  &  2.9      & 180 &      946  \\  % 1.421574e-02
&&&&\\

SE relic          &150  &  382 &  321 &   1689 \\
	  	     &  323 &   114  &   315 &  1660  \\
		     &  610   &   50   &  305   & 1604 \\
		     &	3000	& 5.2 &  280 &  1472\\
		     
Sfilament     & 150     &61$^{*}$ & 120 &  631  \\
		     &  323     &  17        &   160  &  841  \\
		     &  610     &   6.5       &  120    &  631 \\
		     & 3000    &   - & -&-\\

&&&&\\
Total north emission  & 150 MHz  & 1110 & 630 & 3314   \\
&&&&\\
N filament  & 150 &  33 &  270 &  1420 \\
\hline

\multicolumn{5}{l}{\scriptsize  Col. 1: Source name, Col. 2: Observing frequency, Col. 3: Flux density integrated above 3$\sigma$   }\\
\multicolumn{5}{l}{\scriptsize  from the low-resolution maps and subtracting the emission of the embedded radio galaxies.  }\\
\multicolumn{5}{l}{\scriptsize Col. 4:  Largest angular scale of the emission; Col 5: Largest linear scale.}\\
\multicolumn{5}{l}{\scriptsize $^{*}$ At low resolution the emission blends into the one of the nearby source, }\\ 
\multicolumn{5}{l}{\scriptsize  and cannot be unambiguously separated.}\\
\multicolumn{5}{l}{\scriptsize$^{**}$ Only bright spots are detected}\\
\end{tabular}
\end{table*}

\begin{table*}

 \centering
   \caption{Flux density of the radio-galaxies.} %CORRECTED BY TSYS
  \begin{tabular}{c c c c c c c c c c c c c}
   \hline
 Freq. 	&	A    &    B     &    C        &        D   &     E       &   F       &      G     &   H       &     I 	& 	L      &   M         & N    \\
 MHz  &     mJy    &    mJy   &    mJy  &          mJy      &   mJy   &  mJy   &     mJy   &   mJy  &     mJy &   mJy    &    mJy    &    mJy  \\
 \hline
 &&&&&&&&&&&\\
150 &	24   & 16    & 3  &   - &      9  &    15  &    25   &    69 &      44 &     193     &  17 &      21\\
323	&	16.0  &  13.2   &   2.7    &  1.2   &  4.4  &   8.6  &    14.9   &  32    &  28    &  94    &  6.3    &  10.9\\
610 &	10.4  &    9.3  &    2.0  &   1.1    &  2.4   &   6.5    &  10.5    &  19    &  20 &    55   &   3.7  &    10.2\\
3000 &   2.7      &   3.4    &  0.66    &    -          &    0.22 &  0.65       &  1.03     &  5.2   &   4.6 &  6.8     &   -          &         - \\
  
  \hline
\end{tabular}
\label{tab:sources}
\end{table*}  
  
%   Freq. [MHz]		&	A               &            A2             &          B     &         C &                        D   &   			E		         &	F &	F.1 	      &	  F.2	        &		F.3		&G  & H    \\
%150 &	3.382095e-02     & 2.229768e-02   &2.133412e-02   &  3.502931e-02    & 9.911968e-02  &  6.320323e-02 &  2.758480e-01    &9.462749e-03 &	1.239455e-02 &  1.740695e-01 &2.429650e-02  &2.941199e-02\\
%323	&	2.137038e-02	& 1.764784e-02  & 1.149634e-02  & 1.986e-02          & 4.288530e-02  &  3.682106e-02  &   1.248790e-01    &5.302238e-03 &  1.104022e-02    &    8.831465e-02 &8.380837e-03  &1.445914e-02\\
%610 &	1.058060e-02	& 9.444726e-03 	& 6.573810e-03  & 1.061765e-02    & 1.927176e-02  & 2.051639e-02  &     5.616653e-02    &2.724714e-03 &   6.066113e-03   &     4.220137e-02  &3.790181e-03  &1.029353e-02\\

\section{Cluster radio emission}
\label{radioanalysis}
In this section we describe the different radio emitting parts of the cluster PLCKG287.0 +32.9.
Two radio relics and hints of radio halo emission are visible in the NVSS and TGSS \citep{Bagchi11}.
Our observations reveal that the N relic discovered by Bagchi et al. (2011) is actually a radio galaxy, and that another relic is present closer to the cluster center.
A radio halo is definitely established, and additional filamentary emission is detected around the relics.
The wide frequency range covered by our observations allows us to perform a spectral analysis and 
investigate the origin of the observed emission.

\subsection{The northern relic and the filamentary structure}

The cluster is imaged with a sensitivity that is 5-7 times better than the reprocessed TGSS observations analysed by \citet{Bagchi11}, 
and  a resolution that is $\sim$4-8  better than TGSS and NVSS, allowing us to detect additional diffuse emission 
and to analyse the morphology and spectral properties of the different components.
Our observations provide a different view of the radio emission compared to the one given by \citet{Bagchi11}.
The source called northern relic (RN) in \citet{Bagchi11} is actually a radio galaxy, composed of two lobes that are edge-brightened. The emission of the
E lobe is connected to the NW relic (which is called Y-shaped filamentary radio feature in \citealt{Bagchi11}). 
The NW relic dists in projection $\sim$ 350 kpc from the cluster X-ray centre.\\

We detect a low-brightness filamentary structure that departs from the NE tip of the relic and bends towards the W, increasing its
brightness, towards a structure that looks like a radio galaxy with a core and two bent lobes (core, E lobe and W lobe in Fig. \ref{fig:label}).
This emission is not detected at 3 GHz, where the candidate radio galaxy and the relic are disconnected.
This indicates that the electrons producing the relic region come from the radio galaxy (see Sec. \ref{discussion}). However, 
no optical counterpart to the core has been found (Fig. \ref{fig:optradio}).  We discuss this scenario in Sec. \ref{discussion}.\\
At its SW tip, the relic emission becomes less elongated and more roundish. The brightness profile of the relic is symmetric with respect to the major axis of the relic,
 with its maximum in the central part.  Towards the cluster center, the emission merges into the radio halo. In the 3 GHz image, this connection
 between the radio halo and the relic is not detected. A radio bridge connecting a radio halo with a relic has been observed in other clusters (e.g. Coma,  \citealt{1993ApJ...406..399G}).\\
Another radio filament is detected at 150 and 323 MHz close to the SW tip of the NW relic and elongated perpendicular to
the relic's major axis (NW filament). 
At 150 MHz, this filament is attached to the SW part of the relic, while at 323 MHz it is detached.
At 610 MHz and 3 GHz only a small bright spot of the filament is detected.
However, due to the different coverage of the short uv-spacing, it is not possible to draw firm conclusions on the
spectrum  of this region (see discussion about spectral index in Sec. \ref{sec:spix}).
In order to highlight, the full extent of the radio emission, we have imaged the datasets while tapering down the long baselines (Figs. \ref{fig:radio} and \ref{fig:lr}).
At 323 MHz and 150 MHz, more diffuse emission is detected beyond the relic, towards the cluster outskirts. At 150 MHz, this emission encompasses all the different 
structures detected at higher resolution. The total extent of the northern emission, considered as a whole, is $\sim$10$'$, corresponding to a linear scale of $\sim$ 3.3 Mpc at the cluster's redshift.
Such a large extension is not detected in the other bands because of the reduced sensitivity to the large angular size
compared to the 150 MHz observations, combined with the spectral properties of the emission. Another filamentary source, 1 Mpc wide, is detected at 150 MHz north of the radio-galaxy (see Fig. \ref{fig:lr}).\\
Flux densities, dimensions and other properties of the radio emission around the NW relic are listed in Table \ref{tab:radioProp} and labelled in Fig. \ref{fig:label}.
The flux of the extended components has been measured from the low-resolution images after subtracting the contribution of the embedded sources
 (Table \ref{tab:sources}). However, at 150 MHz the different components
are merged into a single patch of emission, and there is no unique way to define the borders of the relic, filaments and lobes. 
Hence, we also measure the total flux of the emission in the NW (relic, filaments, lobes and additional diffuse emission detected NW of the relic).

\subsection{The radio halo and the southern tail}

Our observations confirm the presence of a radio halo, as suggested by \citet{Bagchi11}. The halo emission merges
into the northern relic and has a roundish shape.
At 323 MHz, the radio halo emission roughly follows the X-ray emission of the cluster, as found in other cases \citep[see review by][]{Feretti12}.
At 150 MHz, the low-resolution image reveals a southern extension of the radio halo (Fig. \ref{fig:lr}).
This southern component has no counterpart in the X-ray image of the cluster, has a lower surface brightness with respect to the radio halo,
and a ``double tail" shape, merging into two radio galaxies. 
The total extent of the radio emission is about 450$''$, corresponding to 2.3 Mpc at the cluster redshift. Excluding
the southern component, the roundish halo has a total extent of $\sim 250''$ (1.3 Mpc). Further details are listed in Table
\ref{tab:radioProp}.

\subsection{The southern radio relic and the southern filament}

A second relic is located at a projected distance of 2.8 Mpc from the X-ray center, south of the subgroup detected through the optical analysis
(Sec. \ref{sec:optical}). The SE relic is elongated in the direction NE-SW parallel to the NW relic and perpendicular to the
mass distribution as traced by the galaxies (see Fig. \ref{fig:isodensity}). It has a classical ``arc-like" morphology detected in 
other relics \citep[e.g][]{1997MNRAS.290..577R,2006Sci...314..791B,2010Sci...330..347V}.
As noted by \citet{Bagchi11} the SE relic appears (in projection)  much farther from the cluster center, as detected from X-ray emission, than the NW relic.
Optical data reveals  a group of  galaxies south of the main cluster.
This southern clump is likely a sub cluster, not visible in the X-ray image, which is interacting with the main cluster. 
This situation could be similar to the one
observed in the Coma cluster, where a relic is located behind a sub-group interacting with the main cluster. Alternatively, the galaxies in the southern clump
could be in the process of accreting towards the main cluster through a cosmic filament with little hot gas. \\
New emission is detected in the region of the SE relic (SE filament in Fig. \ref{fig:label}). It has a filamentary morphology
and it is elongated almost perpendicular to the SE relic major axis, with a maximum extension of  $\sim$ 850 kpc towards south at 323 MHz.
The relic is barely visible in the 3 GHz image because of the primary beam attenuation. 
Contrary to the radio halo and to the NW emission, no further emission is detected ahead of the SE relic in the lower frequency images.

\begin{figure*}
\vspace{100pt}
\begin{picture}(100,100)
 \put(210,0.4){ \includegraphics[width=10cm]{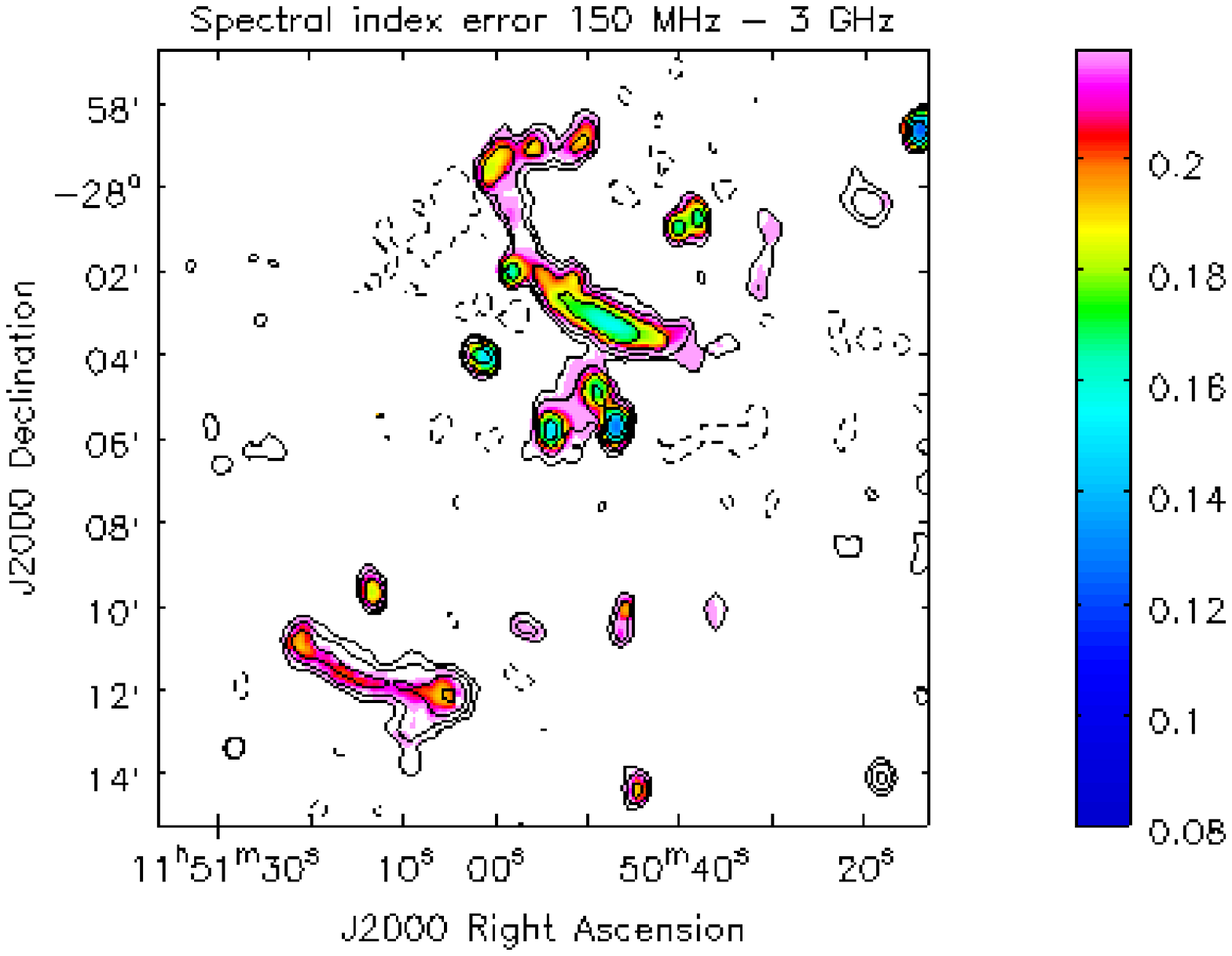}}
 \put(-31,0.4){ \includegraphics[width=10cm]{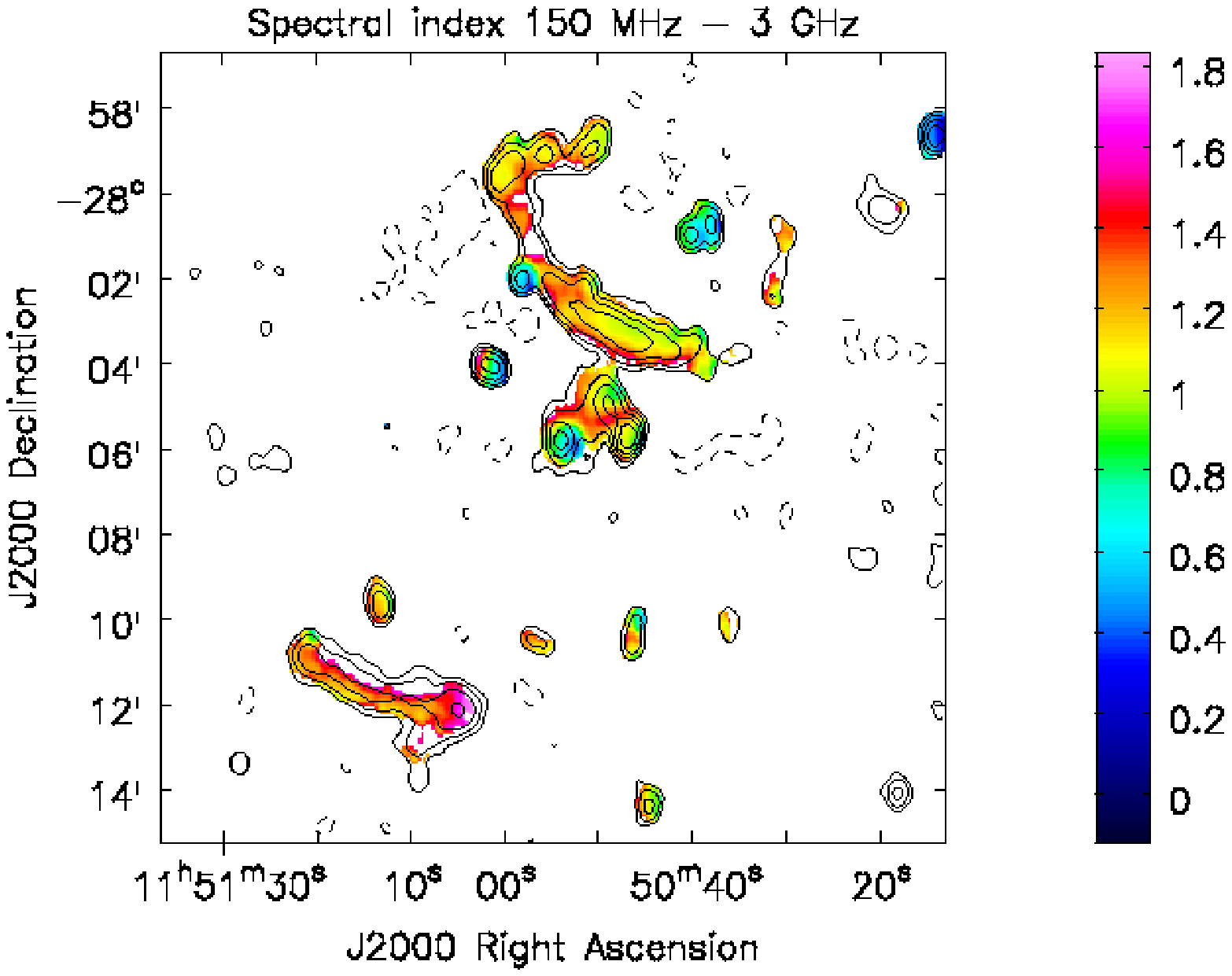}}

\end{picture}

 \caption{Spectral index and spectral index error image obtained considering 150, 323, 610, and 3000 MHz images (colours). 150 MHz contours are overlaid. Note that much of the extended emission is not shown because of the common uv-range
  used,  required to minimise instrumental effects.}
  \label{fig:spix4}
\end{figure*}

\begin{figure*}
\vspace{100pt}
\begin{picture}(100,100)
 \put(210,0.4){  \includegraphics[width=9.8cm]{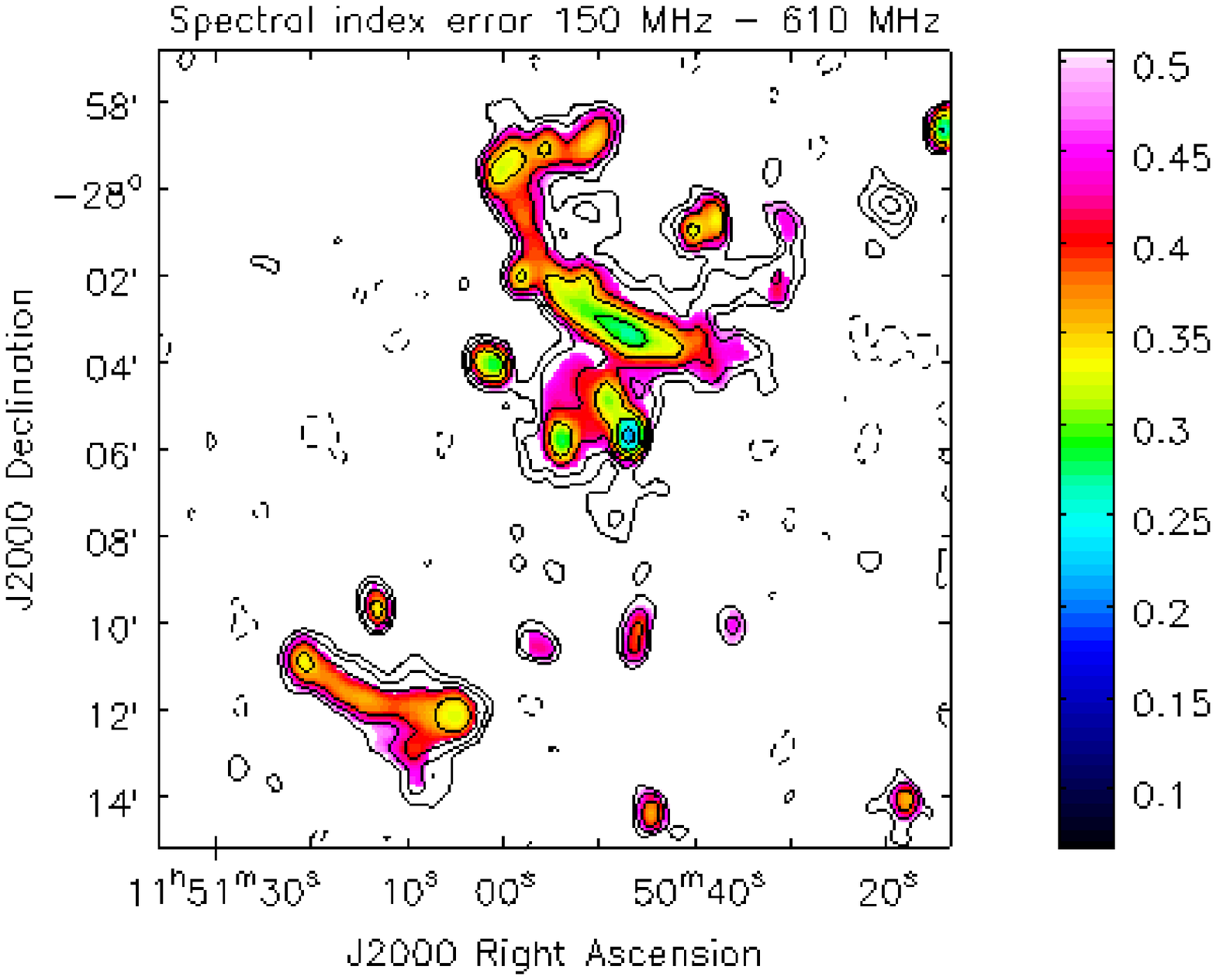}}
  \put(-31,0.4){  \includegraphics[width=9.8cm]{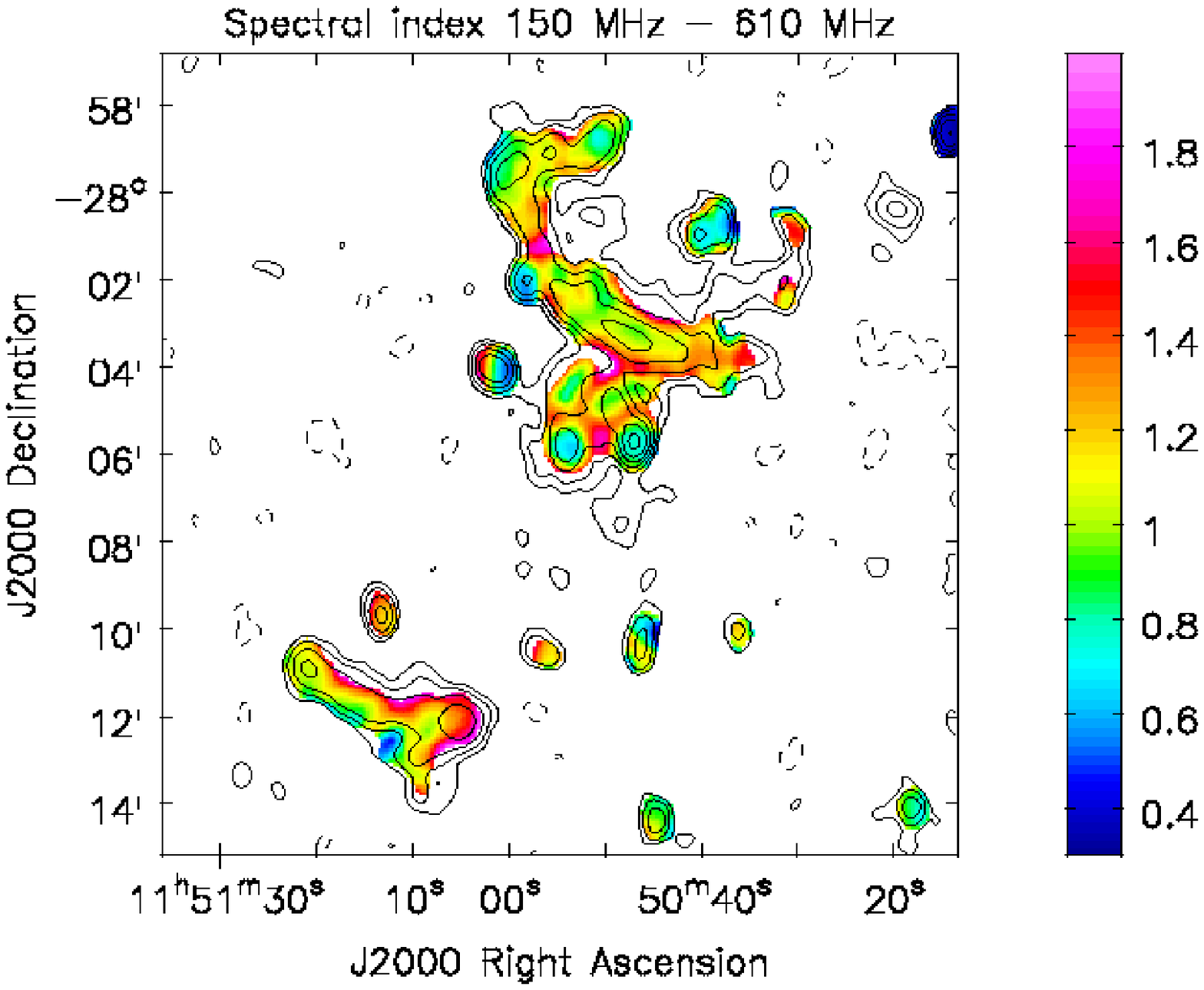}}
\end{picture}
 \caption{Spectral index and spectral index error image obtained considering 150 323 and 610 MHz images (colours). 150 MHz contours are overlaid.}
\label{fig:spixAll}
\end{figure*}

\begin{figure*}
\vspace{100pt}
\begin{picture}(100,100)
 \put(210,0.4){ \includegraphics[width=9.6cm]{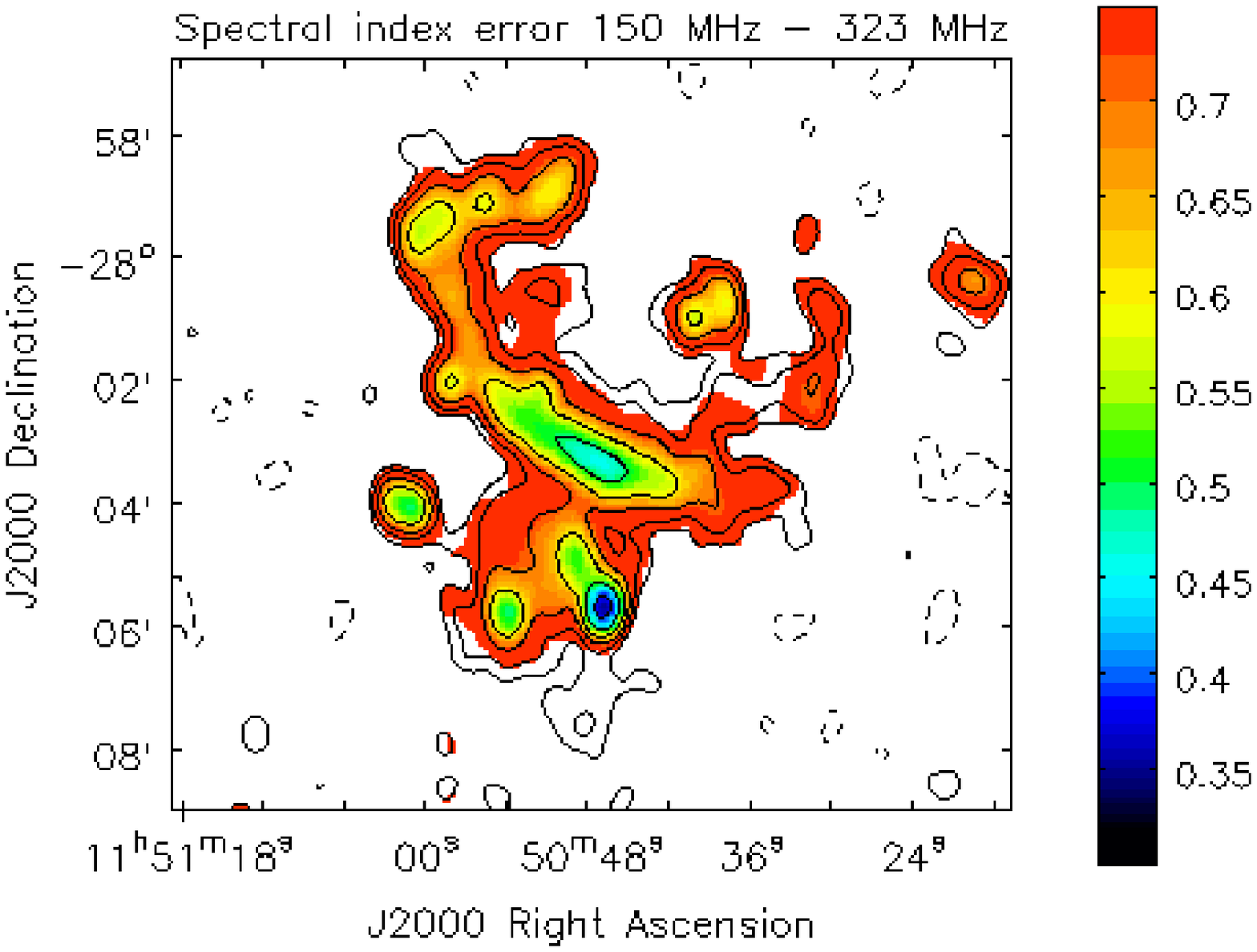}}
  \put(-31,0.4){ \includegraphics[width=9.6cm]{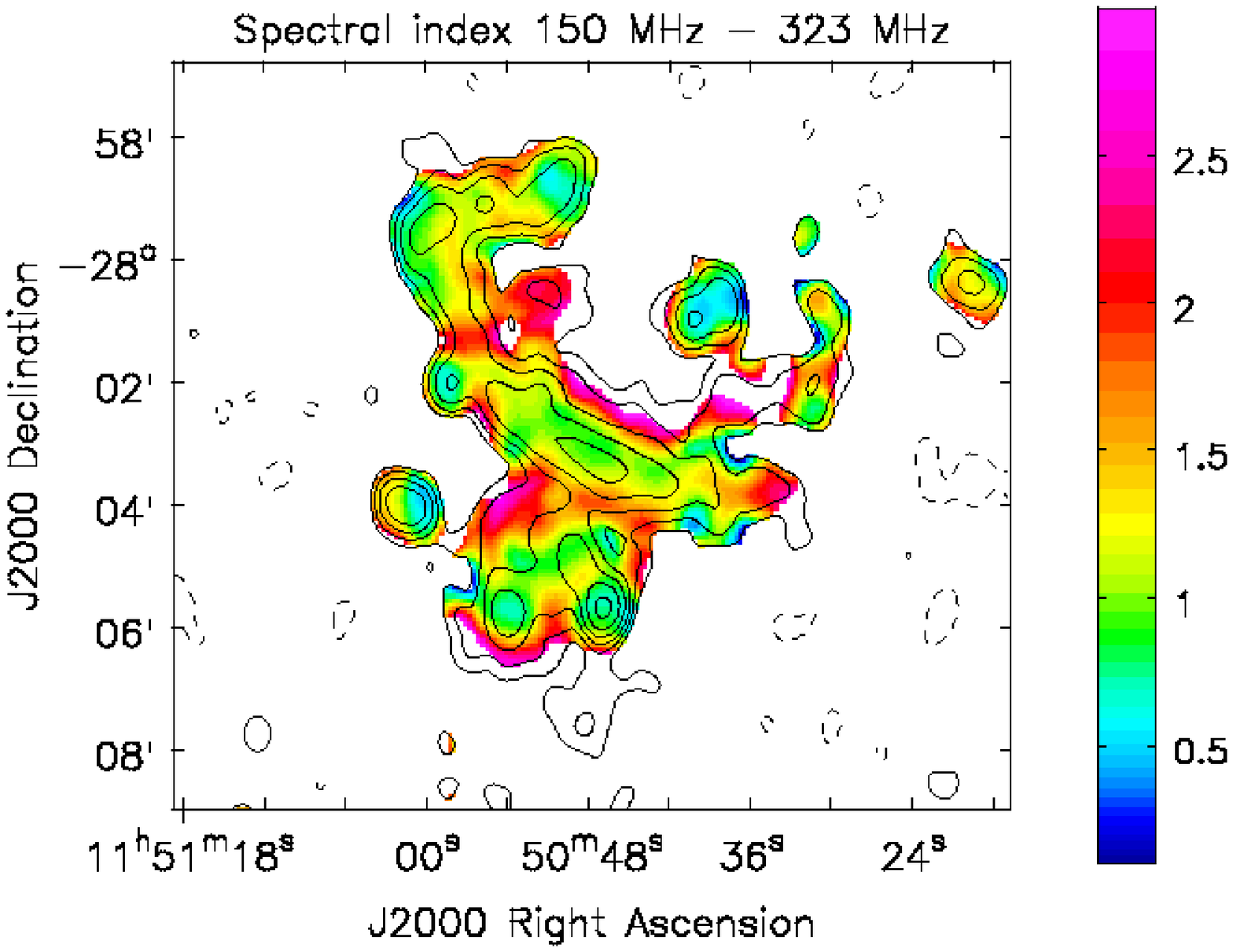}}
  \end{picture}
 \caption{Spectral index and spectral index error image obtained considering 150 and 323 MHz images (colours). 150 MHz contours are overlaid. }
 \label{fig:spixN}
\end{figure*}

\begin{figure*}
\vspace{100pt}
\begin{picture}(100,100)
\put(-31,0){\includegraphics[width=9cm]{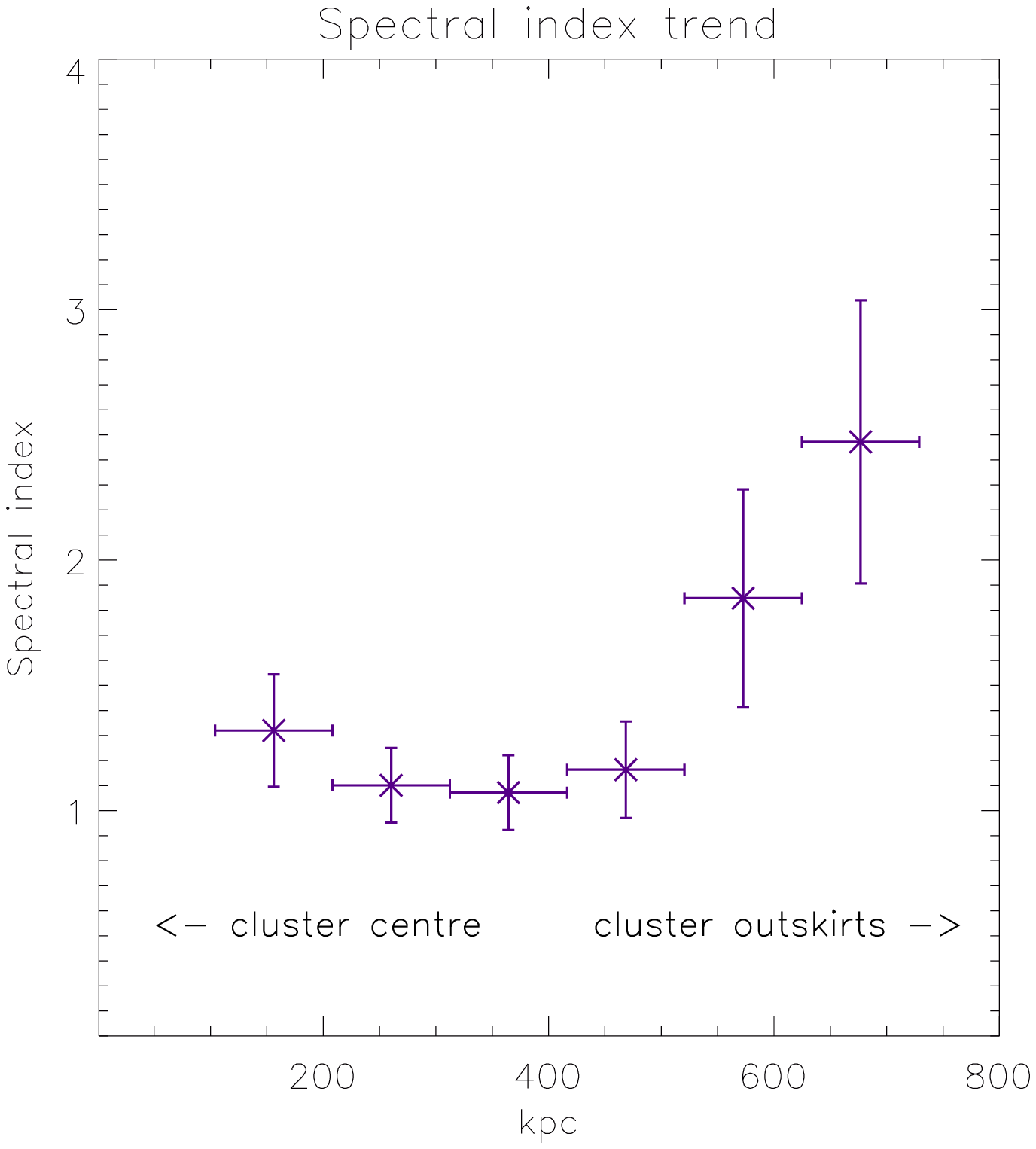}}
\put(18,160){\includegraphics[width=3.5cm]{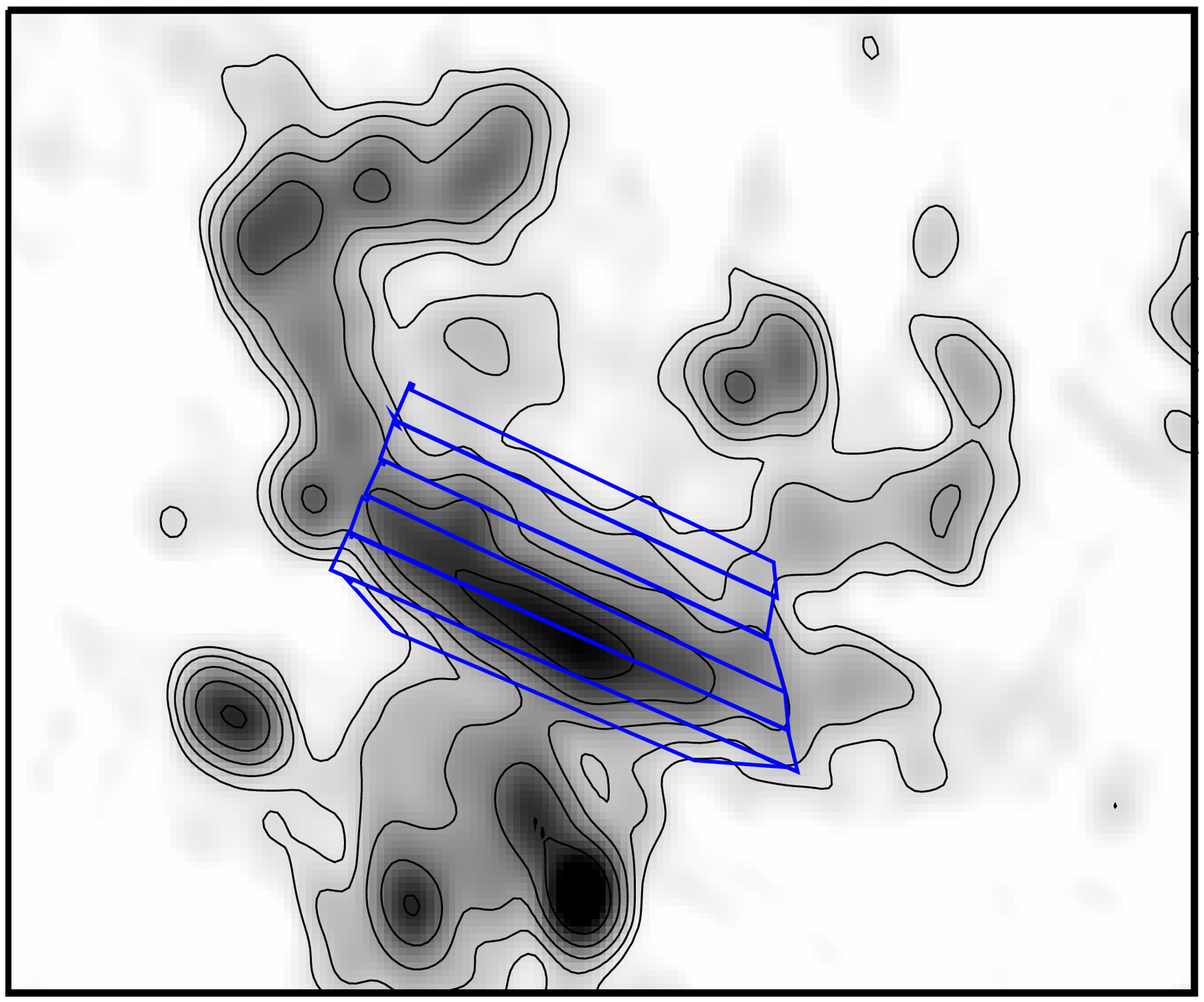}}

\put(230,0){\includegraphics[width=9cm]{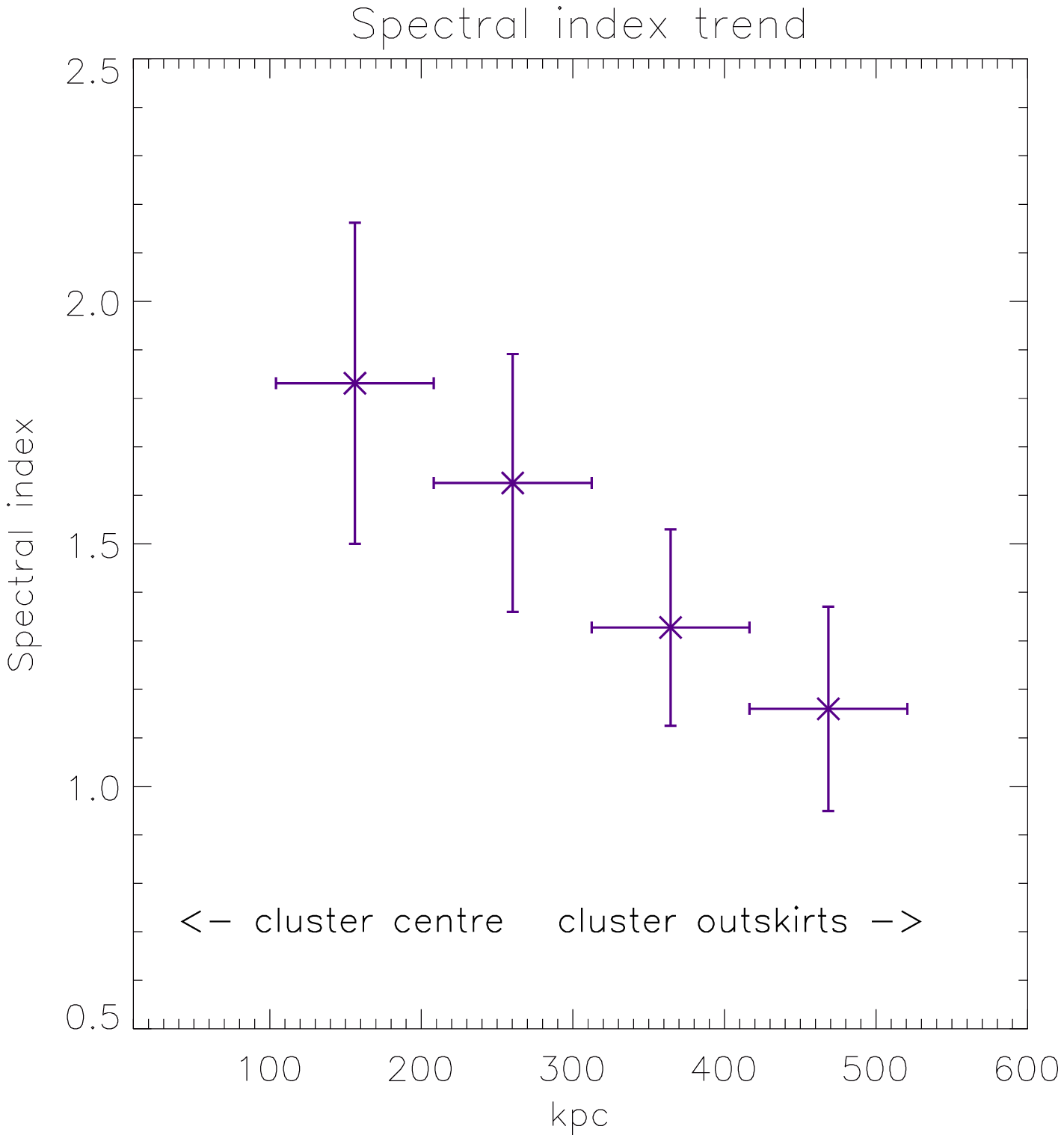}}
\put(380,160){\includegraphics[width=3.3cm]{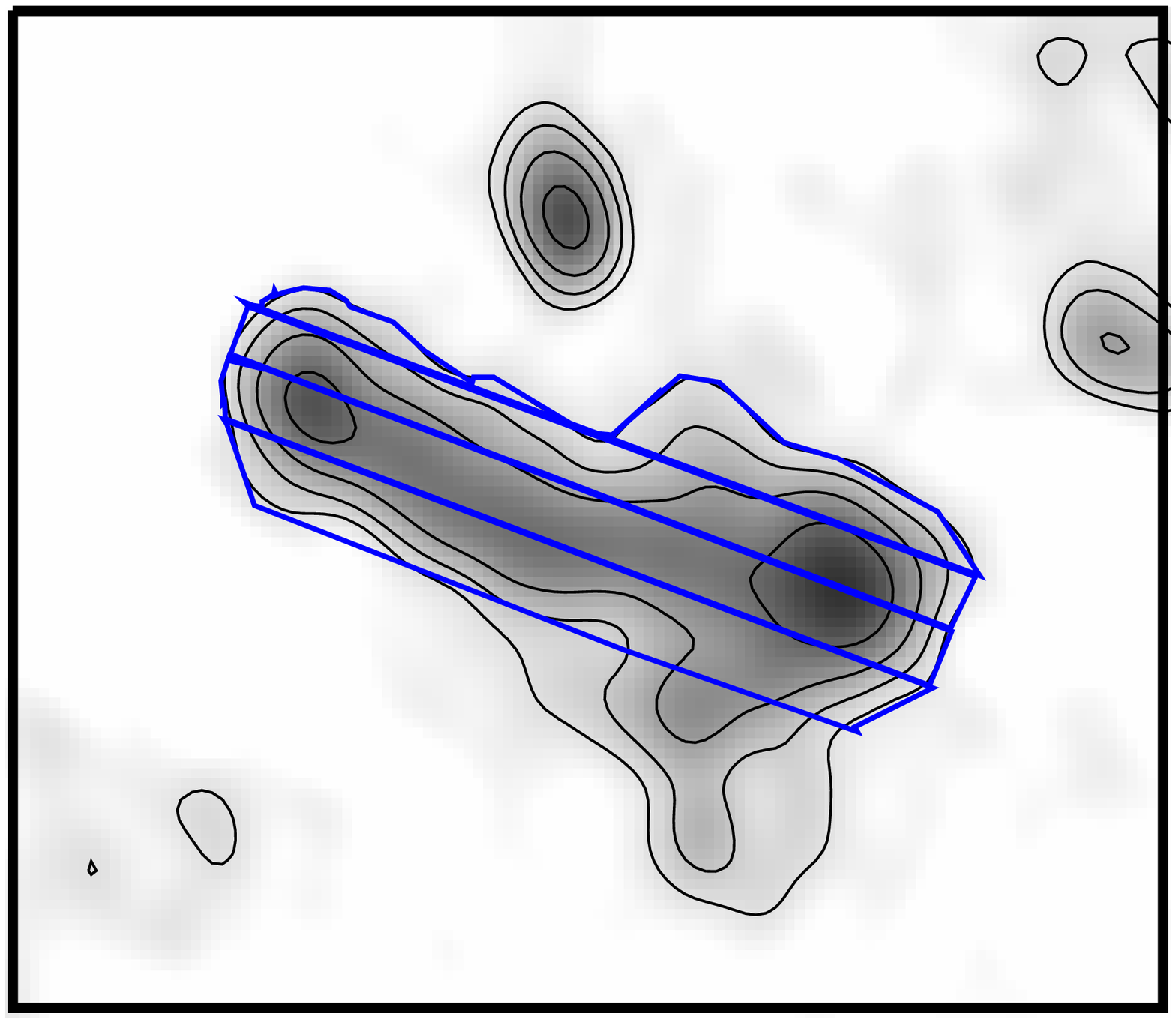}}

\end{picture}
 \caption{Spectral index in the NW (left panel) and SE relic (right panel). The spectral index is computed in the boxes shown in the
 panels. The spectral index has been computed between 150 and 610 MHz.}
 \label{fig:spixreg}
 \end{figure*}

\section{Spectral index study}
\label{sec:spix}
In this Section, we analyse the spectral index maps based on the combined GMRT and JVLA
observations as well as on the GMRT maps only. The latter is relevant to study the spectral properties of the emission that is not detected at high frequency (3 GHz).\\

Using the high-resolution images, we have checked for possible systematic positional offsets in different datasets. These offsets can be generated in the process of self-calibration
or ionospheric-phase corrections. 
Sources detected in the 323 MHz map with a signal-to-noise ratio $>$ 10 have been used to calculate the offsets relative to the NVSS (83 sources were found). 
We have repeated the same procedure for the other images, taking now as a reference the shifted 323 MHz map. We have found 85 sources, 130 sources and 18 sources
in the 150 MHz, 610 MHz and 3 GHz map, respectively. The resulting offsets are small (less than 1$''$), and smaller than the grid resolution used to image the datasets
(4$''$). Nonetheless, we have corrected for them by shifting the visibilities before Fourier-transforming them into the imaging plane.
The datasets have been imaged using the same range of baselines in the 
uv-plane (0.35 - 12.5 k$\lambda$), gridding the data onto the same grid, and tapering the long baselines 
(using a circular Gaussian taper of 8 k$\lambda$) to enhance the diffuse emission. 

After cleaning,  each image has been  corrected  for the primary beam response
and convolved  with the same Gaussian beam (FWHM of the beam 30$'' \times$ 28$''$).
We have then fitted the spectrum using a linear model minimising the chi-square error statistics
wherever the emission was detected above 3$\sigma$ at all frequencies. We note that this introduces a bias towards
the flattest spectral index regions, that can be detected up to 3 GHz.\\
The spectral index image is displayed in Fig. \ref{fig:spix4}. Since the FWHM of the convolution beam is sampled by $\sim$ 64 pixels in the images,
while the fit has been performed pixel by pixel, this image is only meant to show the general features of the spectral index 
distribution. In addition,  the uv-plane distributions remain quite different, even after the uv restrictions that we have applied.
Hence, the spectral index maps have to be treated with caution.  In order to improve the significance of the spectral index variation across the relic, we have integrated the
flux over larger regions (see Sec. \ref{sub:relic1_spix},   \ref{sub:halo_spix},  \ref{sub:relic2_spix}, and Fig. \ref{fig:label}). Embedded sources have been masked.\\
The VLA and GMRT observations have a very different coverage of the uv-plane.
Additional differences come from the different observing time and different sensitivities. For these reasons, it is not possible to derive a spectral index distribution
for the weak diffuse components detected in this cluster using all the observations.\\
We have derived the spectral index distribution at low frequencies by considering the GMRT observations only, to study the spectral properties 
of the emission not detected at high frequency (we imaged the common uv-range 0.12 - 12.5 k$\lambda$  and used a circular taper
as above). The spectral index  image, shown in Fig. \ref{fig:spixAll}, shows more of the diffuse emission.
We have overlaid the 150 MHz contours from which the spectral index map is derived
to point out that even excluding the 3 GHz image, the spectral index could be derived only in certain parts of the image. 
At 150 MHz, additional emission is detected especially around the
NW relic and the radio halo. This emission is less extended at 323 MHz and not visible at 610 MHz above 3$\sigma$. 
Since we have selected the visibilities in the same uv-range, we can conclude that the
synchrotron emission is steep ($\alpha >$ 2) and hence harder to detect at higher frequencies (see however  \ref{warning} for possible 
artefacts  on the spectral index.).

\begin{table*}

 \centering
   \caption{Integrated spectral index in the various regions. }
  \begin{tabular}{c c c c c c}
     \hline
  &W Lobe  &Core&E Lobe & NW relic \\
  $\alpha$ & 1.15 $\pm$  0.17 &  1.20 $\pm$  0.19 & 1.16 $\pm$  0.15  & 1.36 $\pm$  0.21 \\
 &&&&\\
  \hline
   &&&&\\
  &NE filament & Halo & SE relic & NW filament\\
$\alpha$  &  1.16 $\pm$  0.08 & 1.28 $\pm$  0.16 & 1.33 $\pm$  0.11  & 1.21 $\pm$  0.21 \\
  \hline

\end{tabular}
\label{tab:spix}
\end{table*}  

\subsection{The NW relic and northern  emission}
\label{sub:relic1_spix}
The spectral index distribution in this region shows a gradient along the relic's minor axis, a steepening in the NE filament, and a flattening again towards the lobes
of the radio galaxy.  The integrated spectral index in the NW relic is shown in Fig. \ref{fig:label}, and  is well-represented by a single power law from 150 MHz up to 3 GHz.
In order to recover as much information as possible for the extended emission detected NW of the  NW relic, we have created an additional  spectral index map
using only the 150 MHz and 323 MHz observations (Fig. \ref{fig:spixN}). 
Some important trends are visible in the relic. The spectral index shows a gradient along the
relic minor axis, with the flattest part located towards the diffuse emission detected NW of the relic.
The spectrum steepens gradually towards the cluster center, and more sharply  towards the NW diffuse emission in the cluster outskirts up to $\alpha \sim 3.1$. 
 To show this gradient more clearly, we have integrated the flux in the regions displayed in Fig. \ref{fig:spixreg}, and fitted the spectral index in each region separately. 
 Considering only the regions where the flux density is above 3$\sigma$ in all the maps would introduce a bias towards low spectral indexes
 since the emission becomes wider towards lower frequencies. 
 To remove this bias, we have integrated the flux density at 150, 323 and 610 MHz in shells
 parallel to the relic major axis, considering all the pixels where the emission at 150 MHz is above 3$\sigma$. 
  We have fitted the spectral index in each shell, taking into account only the
 flux densities where the flux was above 3 times the error, estimated as $\sigma \times \sqrt{n_{beam}}$ \citep[see e.g.][]{Orru07,Bonafede09a}.
 In the two outer shells, the flux density at 610 MHz  is below this limit, hence only an upper limit can be derived. We have checked that the upper limits
 are consistent with the spectral index obtained using 150 MHz and 323 MHz flux densities.
In Fig. \ref{fig:spixreg} the spectral index in the shells is displayed. The flattest part is in the central shell, located at the edge of the relic
emission, with an average value of $\alpha=1.07 \pm 0.17$. Towards the cluster center, the spectral index steepens gradually with
$\alpha=1.10 \pm 0.17$ and $\alpha=1.31 \pm 0.26$ in the inner shell close to the halo emission. Towards the outskirts of the cluster - where 
diffuse emission is detected only at the lower frequencies - the spectral index steepens with values of
$\alpha=1.15 \pm 0.23$, $\alpha=1.85 \pm 0.55$, and $\alpha=2.47 \pm 0.60$.
If we assume that the particles are being accelerated by a shock that is moving outwards,  in the shell that displays the flattest spectral index the corresponding Mach number $M$ would be $M \sim 5.4$, assuming stationarity and continuous injection. In this case, the trend towards the cluster center can be explained with synchrotron ageing, but the steep emission
detected ahead of the relic is more difficult to explain (see Sec. \ref{discussion}). 
It is worth mentioning that recent simulations of one galaxy cluster \citep{sk13}  are able to reproduce small spectral index variations across radio relics without considering radiative losses. 
They show that the complex shock structure and projection effects are in principle able to reproduce spectral index trends. 
Hence, the steepening in an unexpected direction, NW of the NW relic in our case, does not necessarily imply ageing. 
At the same time, the steepening that we attribute to ageing could be contaminated by projection effects.
The Mach number we derive from the NW relic spectral index would be rare for merger shock waves, although not impossible \citep[e.g.][]{sk08,2011MNRAS.418..960V}. However, radio-derived Mach numbers are notoriously unreliable \citep[e.g.][]{sk08}. 
 The spectral index steepens towards the NE filament, becomes flatter in the E lobe, steepens again towards the core and flattens again 
in the W lobe. The spectral index value in the core is around $\alpha \sim 1$, hence steeper than the one typically detected in the core
of radio galaxies. The trend in the lobes and the flat value of the spectral index detected at the edge of the lobes is typical of FR type II radio galaxies.
In Fig. \ref{fig:label} we show the attempts to fit the emission in the lobes, in the core, in the NE and NW filament with a single power-law (see also Table \ref{tab:spix}).
The spectral index in these regions seem to deviate from a single power law. In the NE filament the spectrum is curved, as well as in the lobes and in the core.
The spectral index in the NW filament, instead, shows a steepening at frequencies $>$ 323 MHz and flattens between 610 MHz and 3 GHz. 
The curved spectra detected in the lobes, core and in the NE filament suggests that the core is not a core of the candidate radio-galaxy, but rather a
distant radio source, and that the plasma emitting in the NE filament was old radio plasma which has undergone adiabatic compression, pushing the synchrotron cut-off to higher frequencies. The optical counterpart could be the small galaxy observed E of the core, although it is likely not a massive elliptical as one would expect. The fact that no jet emission is visible indicates that the radio-galaxy is no more active now, hence a rather faint  radio core
is not surprising.

The spectral index trends detected in these regions suggest that we are witnessing the process of shock re-acceleration. The old plasma is supplied by the lobes of a radio-galaxy
whose nucleus is turned off, and possibly a shock  wave is compressing the plasma in the NE filaments and accelerating the particles in the relic. We will discuss this possibility in more detail in Sec. \ref{discussion}.

 \subsection{The radio halo}
 \label{sub:halo_spix}
 As for the emission around the NW relic, the measured extent of the radio halo becomes smaller going from 150 MHz to 3 GHz. The southern extension is detected only at 150 MHz, 
 indicating that the spectral index must be steeper than $\alpha =$ 3 in that region.
 The spectral index distribution across the radio halo is quite patchy and no clear trend can be detected. 
 If we restrict our analysis to the GMRT observations, the halo is better imaged.
 %Once the sourced are masked, the integrated spectra index in the region shown in Fig. \ref{fig:spixreg} is 1.25 $\pm$  0.20.\\
 In the spectral index maps between 150 MHz and 610 MHz and between 150 MHz and 323 MHz some interesting features are detected.  A patch with a flat spectral index is located in 
 the NE part of the halo ($\alpha_{150 \rm{MHz}}^{610 \rm{MHz}} \sim$ 0.8) which shows a steep spectrum between 150 MHz and 323 MHz ($\alpha_{150  \rm{MHz}}^{323  \rm{MHz}} \sim$ 2.6). This difference is significant although the errors in the spectral index map are large. We have computed the spectrum in this region  separately, deriving an average spectral index of $\alpha_{150  \rm{MHz}}^{323  \rm{MHz}} =1.7 \pm 0.6$ and $\alpha_{332  \rm{MHz}}^{610  \rm{MHz}}=0.3 \pm 0.6$, which within 1$\sigma$ error are incompatible with a power law. The non-detection at 3 GHz is suggesting that the spectrum steepens again above  610 MHz, but due to the missing short baselines no firm conclusions can be drawn. A similar behaviour has been observed in the radio halo of Abell 2256 \citep{vanweeren12}, where a steepening of the spectrum in one small region of the halo has been interpreted as due to  a mixed population of primary and secondary particles, or as an indication of particle re-acceleration by inhomogeneous turbulence.  We have fitted the spectral index between 150 MHz and 3 GHz  (Fig. \ref{fig:label}), restricting to the area of the halo detected at all frequencies. 
 Within the errors, the integrated spectrum is compatible with a single power law.

Estimating the power of the radio halo at 1.4 GHz is not easy. The size of the halo more than doubles between 150 MHz and 3 GHz. 
Since most of the radio halos used to compute the correlation have been observed either at 1.4 GHz or at 300 MHz or 610 MHz, we have extrapolated the flux
from the 610 MHz measurement assuming a spectral index of 1.28.
 The corresponding power of the radio halo at 1.4 GHz, including the K-correction,  would be
 5.1 $\times 10^{24} \, \rm{W Hz}^{-1}$ at 1.4 GHz, which is a factor $\sim$ 4 lower than what expected from the $P_{1.4 GHz} - L_x$ correlation \citep[e.g.][]{Cassano13} and $P_{1.4 GHz} - SZ$ correlation \citep{SommerBasu}.
If we extrapolate the flux from 150 MHz, including the Southern tails of radio halo and assuming a straight spectral index of 1.28, we would obtain a total flux of 17.5 mJy at 1.4 GHz, corresponding to a radio power of 1.0 $\times 10^{25} \, \rm{ W \, Hz}^{-1}$, which is only a factor 2  below the $P_{1.4 GHz} - L_x$ and consistent with the  $P_{1.4 GHz} - SZ$ correlation. This value should be taken as an upper limit to the power at 1.4 GHz, since the spectral index could be steeper than 1.28 in the regions detected only at 150 MHz.

\subsection{The SE relic}
\label{sub:relic2_spix}
The SE relic spectral index shows a steepening towards the cluster center, as expected from DSA models and detected in other cases \citep[e.g.][]{Bonafede09a,2010Sci...330..347V,Giaintucci11,Bonafede12}. The resolution obtained after the uv-range 
cut is not sufficient to properly distinguish the SE radio filament from the relic emission. 
The emission from 150 MHz to 3 GHz can be represented by a single power-law (Fig. \ref{fig:label}), and the average spectral index is steep ($\alpha=1.33 \pm 0.11$).\\
As for the halo and the emission SE  of the SE relic, more diffuse emission is detected at 150 MHz and at 323 MHz. We have computed the spectral index in shells parallel to the relic major axis,
wherever the 150 MHz emission is above 3 $\sigma$,  as we have done for the 
NW relic. The spectral index profile is shown in Fig. \ref{fig:spixreg}. The first shell, closer to the cluster center, exhibits the steepest spectrum, 
$\alpha=1.83 \pm 0.33$ which is computed using only 150 MHz and 323 MHz images, as the emission at 610 MHz falls below 3$\sigma$. We have checked that the upper limit 
derived from the 610 MHz observation is consistent with the derived value of $\alpha$.
The spectral index flattens towards the outskirts of the cluster, up to a value of $\alpha=1.16 \pm 0.22$. 
If we use this value to derive the Mach number of the shock wave responsible for the radio emission (i.e. if we assume that particles are accelerated 
in this region), we obtain $M \sim 3.7$ under the assumptions of stationarity and continuous injection.

\subsection{Imaging artefacts and warnings on spectral index studies}
\label{warning}
Although the efforts in minimising the instrumental differences among the  observations, interferometric imaging can lead to artefacts in spectral index studies, especially
at the edge of extended emissions. The three sources of artefacts are:\\
\noindent(i) Uneven sampling of the short baselines in different observations. Spectral index maps tend to artificially steepen at the edge of extended emission regions mainly because of  under-representation of the extended emission at higher frequencies. Restricting the  imaging to the common range of sampled visibilities in the uv-plane  reduces this problem. However, differences in the uv-plane sampling  still remain,  also in those parts  that are sampled by all datasets.\\
\noindent(ii)  Noise-induced flux boosting in the higher-frequency maps. Spectral index maps can show a flattening at the edge of extended emission regions due to noise-induced flux boosting in the higher-frequency maps. \\
\noindent(iii)  Imaging artefacts. Spectral index maps generated from individual intensity maps at several frequencies are much more sensitive to the way the imaging  is done than the intensity maps themselves. This is because the  spectral index is determined by taking the ratio of maps at different frequencies, which tends to magnify any small variation that was present but hard to recognise in the intensity maps. Such magnified variations will create a patchy spectral index map, which may be what we see in the NW relic region in the spectral index maps (Figs. \ref{fig:spix4},  \ref{fig:spixAll}, and \ref{fig:spixN}).\\
 Systematic errors like these do not cancel out by averaging the fluxes over large areas along the edges. Hence, the trends shown in Fig. \ref{fig:spixreg} at the edge of the emission
could well suffer from the same uncertainties. However, the trends that we detected in Fig. \ref{fig:spixreg} do not depend sensitively on the spectral index computed in the shell at the edge of the emission.
Although not a proof, the fact that apart from  the region NW of the NW relic, the other diffuse sources do not show steep spectra at their edges, 
seems to suggest that despite  the large uncertainties we are picking up the right trends in the NW relic too.
New imaging reconstruction algorithms are being developed \citep[e.g.][]{2009MNRAS.395.1733W,2013arXiv1307.4370C,2013arXiv1311.5282J} and will shed light on these issues in the next future.\\

\begin{figure*}
\vspace{100pt}
\begin{picture}(100,100)
\put(-70,0){\includegraphics[width=13cm]{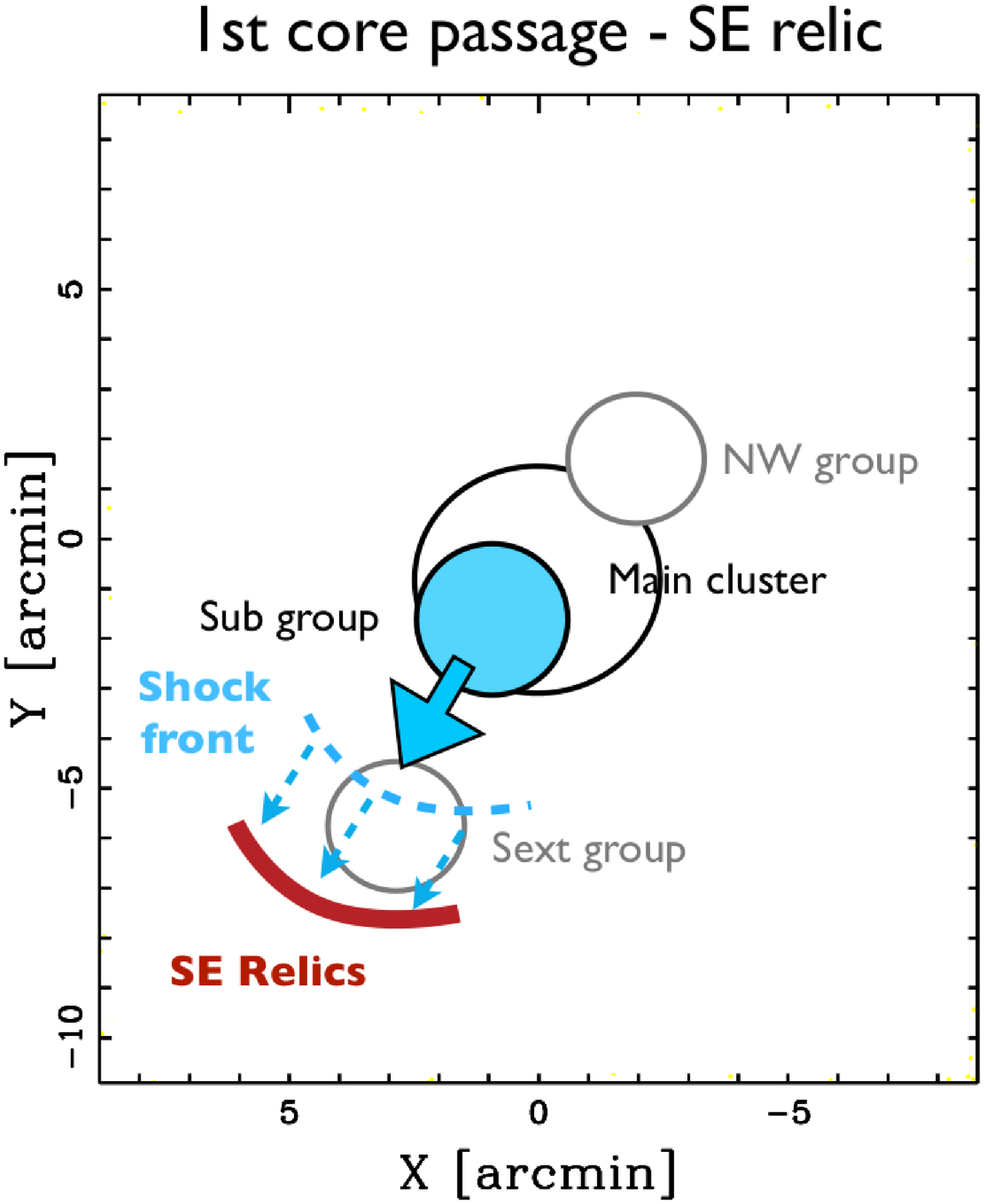}}
\put(190,0){\includegraphics[width=13cm]{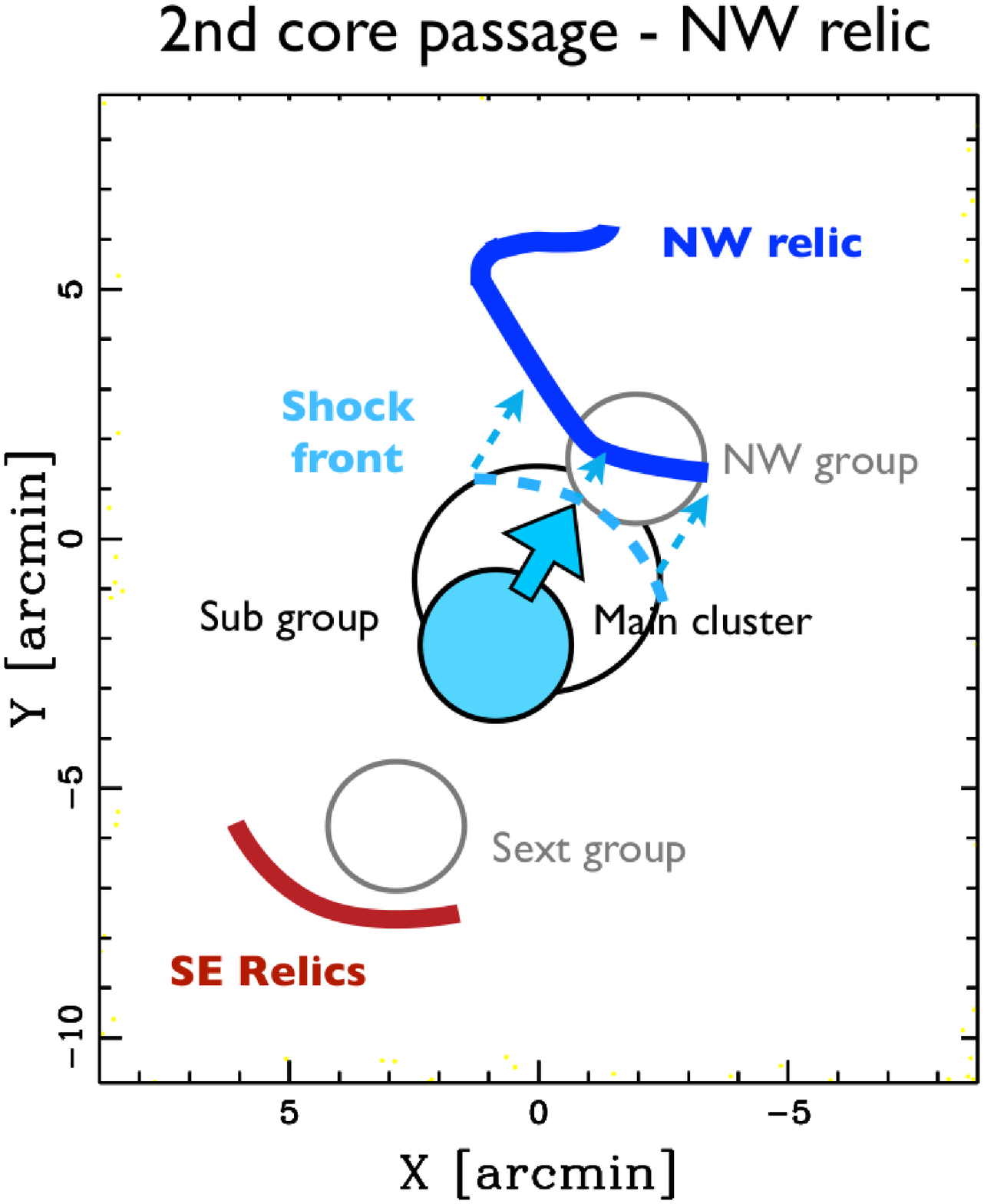}}
\end{picture}
 \caption{  Schematic representation of a possible  merger scenario (case i in Sec. \ref{sec:dyn}). The cyan arrows indicate the
 direction of the motion of the sub-group, dashed lines represent the shock fronts, dashed arrows the direction of the shock fronts that will
 possibly create the relics. Left panel: First core passage and creation of the
 shock wave that generated the SE relic. Right panel: second core passage and generation of the shock front responsible for the NW relic.}
 \label{fig:dynamics}
 \end{figure*}

\section{Discussion}
\label{discussion}

\subsection{Dynamics of the merger}
\label{sec:dyn}
From the analysis of the optical, X-ray, and radio data we can infer some details about the cluster structure
and the merger geometry. The optical analysis shows that the cluster is elongated, likely along an intergalactic filament. In the plane of the sky, it is 20$'$ (6 Mpc) wide and appears to stretch from the SE to the NW. The cluster is undergoing a major merger (SEc - NWc) that is misaligned with the direction of accretion (see Fig \ref{fig:isodensity}). The elongated X-ray emission suggests a past merger between the SEc and NEc 
sub-clusters. The NW relic is located at a projected  distance of $\sim$ 400 kpc from the cluster X-ray center, while
the SE relic is at $\sim$ 2.8 Mpc, which is almost the virial radius for a cluster of this mass.

% Numerical and cosmological simulations suggest that a merger onto the plane of the sky gives the best orientation to detect 
%symmetric shock waves propagating towards the cluster outskirts \citep{2003ApJ...593..599R,va09shocks,2011ApJ...735...96S}. Hence, the presence of double relics has often been claimed as a proof 
%that the merger is observed on the plane of the sky, or close to it.

The different projected distances of the relics from the cluster center
indicate that the shock waves are asymmetrical. We can offer two possible explanations for the origin of radio relics:\\
(i) a sub-group was accreting along the NW-SE direction, along the intergalactic filament, generating a first shock wave at its first
passage through the cluster (SE relic) and a second shock wave (NW relic) on its second core passage (see Fig. \ref{fig:dynamics} for a schematic representation of this merger scenario). In this scenario, the mass
of the sub-group has to be small, otherwise the shocks would have appeared much more symmetrical about the X-ray center. The remnant of the group could be identified in the NW group. The differences in cluster-centric distances could be used to infer the time between the core passages.
A Mach number of $\sim$ 3.7 corresponds to a velocity of $\sim$3700 km s$^{-1}$ in the ICM.
If the shock velocity has remained constant, we estimate that the shock took 0.7 Gyr
to travel from the cluster centre to its position. Similarly, we derive a time of 0.07 Gyr for the shock responsible
for the NW relic emission. If they are both caused by the same subgroup at two different core passages
the time between the core passages is $\sim$0.6  Gyr.
This is just a rough estimate since shock waves change their Mach number while travelling through the ICM,
and the exact path they have travelled is difficult to model. To complicate matters, the small projected distance ($\sim$ 400 kpc) of the NW relic suggests that projection effects may be important.\\
(ii) the SE relic traces a shock caused by the infall of the SEext group into the main cluster, and the NW relic could be due to a 
past  merger episode (e.g. to the interaction with the NW group). Such a scenario resembles the situation in the nearby Coma cluster, where a group is falling into the cluster, and a radio relic appears to be trailing it (Ogrean et al. 2012). 
 However, the curvature and the position of the SE relic  do not seem to be connected to the SEext group.
Velocity information would be required to shed light on the geometry of the system and to infer the direction of accretion onto the cluster in both directions (SE and NW).\\

\subsection{Origin of the steep-spectrum emission ahead of the NW relic.}

The steep spectrum emission detected NW of the NW relic is difficult to explain. According to DSA models, the shock is located where the radio spectral index is the flattest, and steeper radio emission is found downstream of the shock. Here, we detect steep emission either side of the region with the flattest spectrum.   In order to explain the emission ahead of the relic, one could imagine that the particles accelerated at the shock diffuse ahead of it,
producing synchrotron radiation in the upstream magnetic field (the so-called shock-precursor).
The diffusion length $L$ is $L \sim \frac{D}{v}$ with $D$ being the diffusion coefficient and $v$ the velocity of
the particles in the upstream region. Assuming a magnetic field of 1 $\mu$Gauss, for 10 GeV electrons the Bohm diffusion is $\sim 3 \times 10^{23} cm^2 s^{-1}$. For typical velocities
$v \sim 10^{8} cm s^{-1}$, we obtain  $L \sim 0.001$ pc, which is orders of magnitude smaller than the 100s kpc over which the emission is detected.
Alternatively, the shock producing the NW relic could be curved in a direction perpendicular to the plane of the sky with the center of curvature lying NW of the relic. Thus, lines-of-sight that are tangents to the shock surface show the flattest spectrum. Lines-of-sight either side of it pierce regions that contain plasma that has aged and thus yield steeper spectra. This is a somewhat contrived case because it is difficult to explain why an outgoing merger shock wave should assume such a convex shape and why the spectra are steeper towards the cluster outskirts.
We cannot exclude that a complex shock structure observed in projection leads to unexpected trends of the spectral index \citep[e.g.][]{sk13}. 

\subsection{Re-acceleration of fossil radio plasma from AGN}
The complex radio emission detected  around the NW relic suggests that relativistic particles are supplied by neighbouring radio-galaxies. 
In particular, a large radio galaxy (at a redshift of the cluster the radio galaxy would span $>$1.1 Mpc) is detected NE of the NW relic.
 Even though, no optical counterpart has been found, the radio images and spectral index maps suggest that the plasma of the Eastern
  lobe of the radio galaxy is spatially connected to the radio relic. 
The morphology and the spectral features of the NW relic, NE filament and radio-galaxy suggest that the radio-galaxy (lobe E  and lobe W) 
is supplying part of the electrons that are re-accelerated by a shock wave in the relic. However, the lifetime of relativistic electrons is short. 
At  redshift of $z=0.39$, the lifetime is limited by Inverse-Compton losses, 
which limits the lifetime of GeV electrons to less than $10^9$ yrs\footnote{For reference: we follow here Kang, Ryu and Jones (2011)}.
%This time is less than any time needed to spread these electrons along the entire length of the relic. 
%Hence, a supply of CRs ahead of the shock requires tens of radio galaxies that operate in the upstream region of the shock. 
 Electrons with Lorentz factors $\gamma_{L} < 10^2$) have radiative lifetimes larger than the Hubble time.
Within an assumed maximum AGN lifetime of $\sim$ 1 Gyr and a velocity (either a diffusion or bulk velocity) of $\sim 10^8$ cm s$^{-1}$, these low-energy electrons
cannot travel more than $\sim$ 1 Mpc in the ICM.\\
In conclusion, the NE radio galaxy cannot be responsible for the electrons along the entire NW relic. We have to assume that at  least a couple of other radio galaxies have injected CR electrons into the pre-shock region.\\
 Under certain conditions, the spectral features of the radio emission will be determined by the particles injected by the shock: following Kang and Ryu (2011), 
 we assume that the momentum spectrum of a pre-existing population of electrons ($f_{pre}(p)$) can be described by a power law with:
\be
f_{\rm pre}( p) \propto p^{-s},
\ee
and that the shock is injecting particles, above the injection momentum $p_{\rm inj}$:
\be
f_{\rm inj}( p) \propto p^{-q}.
\ee
If the Mach number of the shock is $\geq$ 3, and  $s>4$,  the particles are injected at the shock front 
with a power-law that is steeper than that of the pre-existing electron population (Kang and Ryu 2011, Kang et al. 2012). In this case,  the particles in the downstream region will be dominated by 
the injected particles rather than by the pre-existing population of electrons. In the downstream region, at $p>> p_{\rm inj}$ the spectrum will be:
\be
f_{\rm down}( p) \propto p^{-q}. 
\ee
Hence, the spectral features of the radio emission will be set exclusively by the shock acceleration and lose all memory of the spectrum of the seed electrons. In this case, one can imagine that the electrons 
come from the activity of radio galaxies and diffuse into the ICM outskirts.
The difference in the spectral index features of the pre-existing populations would be erased by the  particles injected by the shock.\\
If the electrons are supplied by radio galaxies, several puzzles posed by radio relics could be solved:
\begin{itemize}
\item{Low-Mach number shocks are not efficient in accelerating particles from the thermal pool, but are efficient enough to re-accelerate cosmic rays from past radio galaxy activity.}\\
\item{Relics are not detected in every merging galaxy cluster. This suggests that a shock wave alone is not sufficient to explain the radio emission. It is also required that a pool of pre-accelerated CRs be present at the location of the relic. Along intergalactic  filament the density of galaxies is higher than in other places equally distant from the cluster center. 
Hence, it is reasonable to assume that a larger number of  electrons are injected by AGN.}\\
\item{No gamma-ray emission is detected in clusters with radio relics, challenging the DSA scenario, \citep{VazzaBruggen13}. If the jet composition is leptonic  \citep{2013APh....43..103R,2013MNRAS.tmp.2432Y,2013arXiv1307.6911Z}, no protons would be injected and hence accelerated by shock waves. Although recently Trigo et al (2013) have found evidence for a baryonic composition of the jets, the debate is still open for AGN-jets \citep[e.g.][]{2013APh....43..103R} .}\\
\end{itemize}
We note that the proposed scenario is different from the one proposed by \citet{2001A&A...366...26E}, where the fossil radio plasma is reenergised by adiabatic compression, and it is not yet mixed with the ICM thermal plasma.

\section{Conclusions}
\label{conclusions}
We have presented a multi-wavelength analysis of the galaxy cluster PLCKG287.0 +32.9, that might shed new light on the origin of radio relics. 
Our results can be summarised as follows:
\begin{itemize}
\item{Optical data suggests that PLCKG287.0 +32.9 is located within a  6-Mpc long intergalactic filament. The cluster has undergone a major merger (NWc -SEc), slightly misaligned with respect to the main 
direction of accretion. Along the filament two sub-clumps are detected. The galaxies along the intergalactic
filament, as well as the two sub-clumps (SEext and NW),  are likely on the process of accretion onto the main cluster.}
\item{Two radio relics and a radio halo are present in the cluster. 
Additional emission is detected NW of the relic. Another relic is located SE of the cluster, at a projected distance of  2.8 Mpc. A radio halo is located
in the cluster center, and filamentary emission is detected around both relics.}
\item{The large projected distance of the SE relic and the small projected distance on the NW relic are interpreted as 
shock waves produces at different times during a minor merger event. A sub-cluster (with a mass $\sim$ 0.1 of the main one) 
could have caused a first shock wave at his first infall onto the main cluster, and a second shock wave during the second core-passage (NW relic).
The radio halo could be due to the major merger between the SEc and NEc sub-clusters.}
\item{The NW relic emission fades into the lobes of a radio galaxy, indicating that radio relics originate from
electrons previously injected by AGN and reaccelerated, likely by a shock wave.
The spectral index analysis supports this interpretation.  However, it is necessary to assume that two or more radio sources have injected electrons into
the pre-shock region to account for the relic size.}
\end{itemize}

\begin{acknowledgements}
We thank F. Vazza and  T. Jones for useful discussions.
A.B. and M.B acknowledge support by the research group FOR 1254 funded by the Deutsche Forschungsgemeinschaft:
"Magnetisation of interstellar and intergalactic media:the prospects of low-frequency radio observations".
The National Radio Astronomy Observatory is a facility of the National Science Foundation operated under cooperative 
agreement by Associated Universities, Inc.
We thank the staff of the GMRT that made these observation possible. GMRT is run by the National center for
Astrophysics of the Tata Institute of Fundamental Research.
RJvW is supported by NASA through the Einstein Postdoctoral grant number PF2-130104 awarded by the Chandra 
X-ray Center, which is operated by the Smithsonian Astrophysical Observatory for NASA under contract NAS8-03060.
This research has made use of the NASA/IPAC Extragalactic Data Base 
(NED) which is operated by the JPL, California Institute of Technology, under contract with the National Aeronautics and 
Space Administration.

%-Please notify Eso Library at esolib@eso.org upon acceptance or
%publication of a  paper based on ESO data, including the bibliographic
%reference (article  title, authors, journal title, volume, year, pages).

\end{acknowledgements}

\bibliographystyle{aa}
\bibliography{master}

\end{document}